# A Simulation-Driven Digital Workflow Integrating Bayesian Optimization for Early-Stage Process Design: A Methyl Acetate Production Case Study

Trung Ngoc Trinh[1, †], Trung Thanh Nguyen[2,†], Nikolaos Tsakoumis[1], Roman Tschentscher[1]
[1]*SINTEF Industry, Process Technology 7034 Trondheim, Norway*
[2]*Faculty of Information Technology- Da Nang University of Science and Technology, Vietnam*
Corresponding author
Trung Ngoc Trinh: trung.trinh@sintef.no
[†]*These authors contributed equally to this work.*

# Abstract

Methyl acetate production via direct carbonylation of dimethyl ether and carbon monoxide involves strong nonlinear interactions among reactor operation, recycle structure, and energy consumption, making conventional simulation-based optimization computationally demanding. In this work, a data-driven framework integrating steady-state process simulation, surrogate modeling, and Bayesian optimization was developed for process intensification. A dataset of 2730 steady-state simulations was generated using a Python-based automated process simulation workflow and a structured design of experiments across the operating domain. Surrogate models were trained to predict carbon conversion yield (C-yield) and cost of specific energy consumption (C-SEC), with neural network–based models showing the best predictive performance and robustness. These models were integrated into a multi-objective Bayesian optimization framework to identify optimal operating regions. Results indicate that high C-yield (up to ~95%) can be achieved with reduced energy cost through lower recycle flow rates and compression demand. Pareto analysis revealed a nonlinear trade-off between C-yield and C-SEC, where further improvements near the conversion limit require disproportionately higher energy input. Validation against rigorous simulations showed deviations below 3% for C-yield and 10% for C-SEC

# 1. Introduction

Methyl acetate (MeAce) is an important platform chemical that can be further hydrogenated to produce ethanol, a widely used fuel additive and industrial solvent (Ait El Fakir et al., 2025; Turek et al., 1994). Among the various production routes, the direct carbonylation of dimethyl ether (DME) with carbon monoxide has emerged as a promising alternative pathway (Kim et al., 2024; Sardesai et al., 2002). This process enables the efficient conversion of C1 feedstocks into MeAce in a single step without water formation, thereby mitigating corrosion issues and simplifying downstream separation. Furthermore, DME carbonylation is inherently compatible with methanol- and syngas-based value chains, making it an attractive option for integrated and sustainable chemical production.

Despite these advantages, the industrial implementation of DME carbonylation remains challenging due to the strong coupling between reaction performance, recycle structure, and energy consumption (De Falco et al., 2017; Diemer and Luyben, 2010; Kim et al., 2024; Zuo et al., 2014). The process typically involves multiple interacting operating variables, such as reactor temperature, reactor length, system pressure, feed composition, and recycle ratio. Variations in these parameters can simultaneously affect MeAce production, conversion efficiency, and energy demand of compressors, heater and cooler, and separation units. Consequently, identifying optimal operating conditions that maximize MeAce production while minimizing energy consumption constitutes a high-dimensional and nontrivial optimization problem.

From a process intensification perspective, improving MeAce yield and reducing energy demand through systematic operating-window optimization represents a promising and cost-effective strategy (Diemer and Luyben, 2010; Liu et al., 2023). However, conventional trial-and-error approaches or one-factor-at-a-time sensitivity analyses are inadequate for capturing complex nonlinear interactions among operating variables. Moreover, direct optimization based solely on detailed process simulation is computationally expensive, particularly when extensive exploration of the design space is required.

Process modeling and simulation have been widely used to analyze and optimize MeAce production via DME carbonylation (De Falco et al., 2017; Diemer and Luyben, 2010; Kim et al., 2024; Zuo et al., 2014). Previous studies have established steady-state flowsheet models to evaluate reaction–separation behavior and to assess the influence of key operating parameters (Diemer and Luyben, 2010). While these models provide valuable mechanistic insight, their application is often limited to local sensitivity analysis or narrow operating ranges due to the computational cost associated with large-scale parametric studies.

Data-driven methods have gained increasing attention in chemical engineering for surrogate modeling, optimization, and decision support (Chatterjee et al., 2019; Hou and Behdinan, 2022; Zhou and Ren, 2025). Machine learning models have been successfully applied to approximate complex input–output relationships in reaction and separation processes, enabling rapid prediction of process performance across wide operating domains. When combined with advanced optimization techniques such as Bayesian Optimization (BO), surrogate models allow efficient exploration of high-dimensional design spaces while explicitly accounting for uncertainty. This combination significantly reduces the computational burden associated with conventional simulation-based optimization and enables systematic identification of optimal operating conditions.

In this study, a data-driven process intensification framework is proposed for MeAce production via DME carbonylation. The framework integrates steady-state process simulation, large-scale data generation, machine learning surrogate modeling, and Bayesian Optimization to systematically explore and optimize the operating space. The main contributions of this work are summarized as follows:

- A systematic exploration of the high-dimensional operating space of the MeAce process using steady-state simulations, automated via Python scripts to enable large-scale simulation campaigns.
- The development and validation of data-driven surrogate models capable of accurately predicting carbon yield and the cost of specific energy consumption across both in-domain and out-of-domain conditions, based on a large simulation-generated dataset.
- The application of surrogate-assisted Bayesian Optimization for multi-objective process intensification, enabling the identification of optimal operating regions and quantitative trade-offs (Pareto front) between production performance and energy efficiency.

The scope of this work is limited to process-level analysis and intensification. The reaction kinetics, catalyst formulation, and base flowsheet configuration (see figure 1) are adopted from a previous study (Diemer and Luyben, 2010). Instead, the emphasis is placed on exploiting simulation-derived data to construct a robust and computationally efficient decision-support framework. As such, the proposed approach is intended to complement, rather than replace, detailed mechanistic modeling and experimental studies, and to provide actionable insights for process optimization and intensification.

# 2. Methodology

## 2.1 Overall workflow and methodological framework

The methodological framework is designed to support process intensification of MeAce production via DME carbonylation. It integrates systematic exploration, data-driven modeling, and operating-condition optimization. The workflow integrates steady-state process simulation, large-scale data generation, surrogate modeling, and Bayesian optimization into a coherent and modular framework. The optimization focuses on two primary objectives: carbon conversion yield and the cost of specific energy consumption.

- C-yield is defined as the ratio of carbon in the MeAce product to the total carbon in the feed streams (DME and CO).
- C-SEC, expressed as the cost of energy consumption per unit mass of MeAce produced, includes the costs associated with heating and cooling duties, steam consumption in the distillation column, and electrical power required for the compression of both fresh feed streams (CO and DME) and recycle streams. The unit prices of these utilities are based on typical market values in Norway.

The first stage of the workflow consists of a steady-state process simulation of the DME carbonylation process. An established flowsheet configuration from literature (Diemer and Luyben, 2010) is employed as the base process model, allowing the study to focus on operating-window intensification. The process model is implemented in a steady-state simulation environment and configured to enable automated parametric studies across a wide range of operating conditions. This stage provides a physically consistent basis for evaluating process performance and cost of energy consumption under different operating scenarios.

In the next stage, a structured Design of Experiments (DOE) strategy is used to generate a large and representative dataset from the process simulation model. Key operating variables governing reactor temperature, reactor length, splitter ratio of purge gas, flow rate of CO and system pressure are systematically varied within predefined bounds to ensure sufficient coverage of the high-dimensional operating space. The resulting simulation campaign yields a large set of converged steady-state cases, from which relevant process performance indicators, such as C-yield and C-SEC, are extracted. Data quality and coverage are assessed to ensure suitability for subsequent data-driven modeling.

Subsequently, stage focuses on the development of data-driven surrogate models. Machine learning techniques are employed to establish input–output relationships between operating variables and key performance indicators. Separate surrogate models are constructed for C-yield and C-SEC. To ensure robustness and generalization capability, the models are trained and evaluated using a structured validation protocol that includes both in-domain and out-of-domain operating conditions. These surrogate models provide a computationally efficient approximation of the underlying process simulation, enabling rapid evaluation of process performance across the operating space.

Finally, the surrogate models are integrated into a Bayesian optimization framework to perform process intensification studies. The optimization framework is formulated to address multi-objective problems, targeting maximization of C-yield and simultaneous minimization of C-SEC. By leveraging the predictive capability of the surrogate models, the Bayesian optimization strategy enables efficient exploration of the high-dimensional design space while balancing exploration and exploitation. This approach allows identification of optimal operating regions and quantitative trade-offs between competing objectives with a significantly reduced computational cost compared to direct simulation-based optimization.

Overall, the proposed workflow provides a systematic and scalable approach for process intensification studies based on simulation-derived data. By combining steady-state process simulation with data-driven surrogate modeling and advanced optimization techniques, the framework enables efficient identification of optimal operating conditions and supports informed Data-driven decision-making for improving process performance and energy efficiency.

## 2.2 Process description and simulation model

### 2.2.1. Process overview and reaction system

The process simulation of MeAce production via direct gas-phase carbonylation of dimethyl ether (DME) with carbon monoxide (CO), as shown in Figure 1, is based on the literature model (Diemer and Luyben, 2010) for process design and life-cycle assessment studies. The process is configured as a continuous reaction–separation system comprising feed compression and conditioning, catalytic reaction, phase separation, distillation, and gas recycle. Fresh DME and CO are pressurized and combined with recycled gas streams before entering the reactor section. Carbonylation occurs in a gas-phase reactor operated at elevated temperature and pressure, producing MeAce together with unreacted reactants and light gaseous components. The reactor effluent is subsequently cooled and sent to a separation unit, where condensable products are separated from the gas phase. The liquid stream is routed to a distillation column for purification of MeAce, while the overhead gas stream is partially purged to prevent inert accumulation and partially recycled to the reactor inlet. The recycle loop enhances reactant utilization but simultaneously increases gas circulation and compression requirements. As a result, overall process performance is governed by the interplay between reactor operating conditions, recycling intensity, and separation efficiency. This configuration leads to a strong coupling between MeAce production, conversion efficiency, and energy consumption, making the identification of favorable operating windows a key challenge for process intensification.

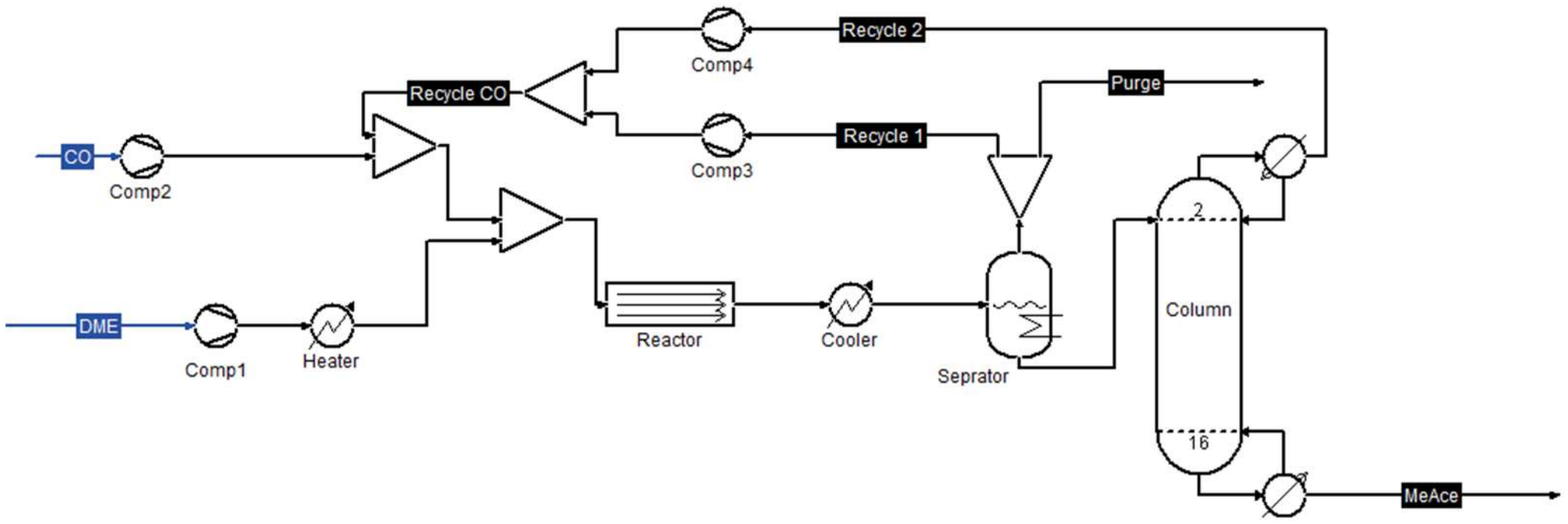

Figure 1: CoCo Process simulation for MeAce from CO and DME adopted from a literature (Diemer and Luyben, 2010)

### 2.2.2. Steady-state simulation setup

The steady-state process simulations in this study were carried out using the COCO (CAPE-OPEN to CAPE-OPEN) process simulation environment ("cocosimulator.org," 2026). COCO is an open-source, CAPE-OPEN–compliant simulation platform that enables integration of thermodynamic packages, unit operation models, and flowsheet solvers within a modular and extensible framework. Owing to its open architecture and scripting capability, COCO is particularly well suited for systematic parametric studies, automation, and large-scale data generation for process analysis and optimization. In addition, COCO allows seamless coupling with external programming environments, such as Python, through standardized interfaces, enabling automated execution of steady-state simulations.

The kinetic models for the production of MeAce from DME and CO have been extensively investigated in the literature (Cheng et al., 2017; Cheung et al., 2006; Diemer and Luyben, 2010; Kipnis and Volnina, 2022; Sardesai et al., 2002). These kinetic expressions have also been incorporated into whole-process simulations in several studies (Diemer and Luyben, 2010; Kim et al., 2024), primarily for process design, techno-economic assessment, and life-cycle assessment. In this work, the base process configuration, reaction scheme, and kinetic description are reproduced from a previously published study (Diemer and Luyben, 2010), which has been made publicly available on the CoCo process simulation homepage by ChemSep ("cocosimulator.org," 2026).

This steady-state formulation is appropriate for the objectives of the study, which focuses on evaluating process performance and cost of energy demand as functions of operating conditions rather than dynamic behavior or control performance. The simulation model is configured to support automated parametric studies over a wide operating range, with key unit operations represented using established modeling approaches consistent with the flowsheet. To enable large-scale data generation, the simulation environment is interfaced with external python scripts for automated control of input variables, execution of simulation runs, and extraction of converged results. Only physically feasible and numerically converged steady-state solutions are retained for subsequent analysis, ensuring consistency of the generated dataset.

Based on the simulation setup, five operating variables are selected as decision variables for process intensification analysis: reactor length, reactor temperature, system pressure, CO molar flow rate, and purge gas splitter ratio. These variables represent the primary degrees of freedom governing reaction severity, recycle behavior, and overall energy consumption. Their operating ranges are defined to ensure physical feasibility and industrial relevance. The carbonylation of DME with CO to form MeAce is a strongly exothermic reaction. The heat released from the reaction is internally integrated and utilized for preheating the incoming DME feed stream. A pinch analysis approach is applied to estimate the overall heating and cooling requirements of the process (the detailed results of the pinch analysis are not explicitly discussed in this study). To avoid excessive model complexity and to ensure clear interpretation of the optimization results, the configuration of the MeAce separation column is kept fixed throughout the simulations. Based on these considerations, the present study focuses on reactor-level optimization with two primary objectives: (i) maximization of C-yield and (ii) minimization of C-SEC.

### 2.2.3. Automation of CoCo process simulation using Python

The process simulation environment employed in this study is primarily designed for interactive execution of individual steady-state simulations. However, systematic exploration of high-dimensional operating spaces and generation of large datasets required for data analytics, surrogate modeling, and optimization necessitate automated execution of a large number of simulation runs. Manual operation under such conditions would be impractical and error-prone.

To address this requirement, the process simulation was first unzipped and converted into an XML-based format, enabling direct interaction with Python for automated control and data exchange. The simulation software was interfaced with Python via a Win32-based communication interface, allowing external scripts to programmatically modify input variables, execute simulation runs, and retrieve simulation outputs. Python scripts were developed to generate operating conditions according to the DOE structure, assign the corresponding input parameters to the simulation model, trigger steady-state calculations, and extract converged results.

Following each simulation run, operating conditions for the subsequent simulation were programmatically assigned from Python to the COCO simulation environment. Instead of exporting results after every individual run, simulation outputs were collected from all converged FSD files (CoCo process simulation format), which follow the standard format of the COCO process simulation, after completion of the entire simulation campaign. The extracted process variables and simulation results were then aggregated and stored in a CSV file. This workflow ensures consistent execution of sequential simulation runs, systematic post-processing of simulation results, and efficient construction of large datasets suitable for subsequent data-driven analysis.

Using this automation framework, 2730 simulation runs were executed, from which converged and physically feasible cases were retained for analysis. The resulting datasets were stored in CSV format to facilitate transparency and seamless integration with data analytics, machine learning, and optimization workflows. The automation of process simulation thus serves as a critical enabling step in the proposed methodology, bridging detailed steady-state process modeling and large-scale data-driven analysis while preserving physical consistency of the underlying process model.

### 2.3. Design of Experiments and data generation

#### 2.3.1. Design of Experiments strategy

To systematically explore the high-dimensional operating space of the DME carbonylation process, a structured Design of Experiments (DOE) strategy was employed. The objective of the DOE was to achieve sufficient coverage of the operating domain while maintaining a manageable number of simulation runs.

**Table 1**: DOE structure for five parameters

| Parameter | Unit | Level 1 | Level 2 | Level 3 | Level 4 | Level 5 | Level 6 | Level 7 |
|---|---|---|---|---|---|---|---|---|
| Reactor length | m | 4.2 | 4.9 | 5.7 | 6.5 | 7.2 | 8.0 | 8.8 |
| Reactor temperature | K | 433.0 | 441.8 | 450.6 | 459.5 | 468.3 | 477.1 | 486.0 |
| System pressure | MPa | 2.10 | 2.35 | 2.60 | 2.85 | 3.10 | 3.35 | 3.60 |
| CO molar flow rate | $mol\cdot s^{-1}$ | 20.0 | 28.3 | 36.7 | 45.0 | 53.3 | 61.7 | 70.0 |
| Splitter ratio of purge gas | | 0.05 | 0.10 | 0.15 | 0.20 | 0.25 | 0.30 | 0.35 |

A Latin Hypercube Sampling (LHS) approach (M. D. McKay, R. J. Beckman and W. J. Conover, 1979.) was selected due to its ability to generate space-filling designs in high-dimensional continuous spaces. Compared to full-factorial designs, LHS provides a more efficient sampling strategy by reducing the number of required experiments while preserving statistical representativeness. The initial sampling was performed in a normalized continuous space, after which the sampled points were discretized into predefined operating levels to ensure physical interpretability and consistency with industrial operating constraints.

The primary focus is placed on optimization of the MeAce reactor, while avoiding an excessively complex system involving a large number of interacting parameters that could compromise controllability. Five operating variables were considered in the DOE, each defined within a feasible operating range and discretized into seven evenly spaced levels (see table 1). The selected ranges were chosen to cover typical operating conditions while avoiding regions associated with numerical instability or physically unrealistic behavior. This approach enables systematic exploration of the combined effects of multiple operating variables on process performance and energy demand.

### 2.3.2. Simulation campaign and data construction

The DOE sampling points were evaluated using the steady-state process simulation model described in Section 2.2.2. Each simulation run was executed under steady-state conditions, and the resulting process outputs were recorded once convergence was achieved. Simulation cases that failed to converge or resulted in physically infeasible operating conditions were excluded from further analysis.

In total, approximately 2730 converged steady-state simulation cases were generated, providing a comprehensive and space-filling representation of the operating domain. For each simulation case, the values of the operating variables and the corresponding performance indicators, C-yield and C-SEC, were extracted and stored in a structured data format suitable for subsequent data-driven analysis, surrogate modeling, and optimization.

To support different stages of model development, validation, and optimization, the simulation results were organized into four distinct datasets. The explicit separation of datasets enables transparent assessment of surrogate model performance, generalization capability, and optimization reliability.

- Dataset 1 (DOE dataset for model development) consists of 2730 steady-state simulation samples generated from the reduced DOE. This dataset provides a structured and space-filling coverage of the design domain and serves as the primary data source for exploratory data analysis, feature interaction assessment, and training of data-driven surrogate models.
- Dataset 2 (in-domain verification dataset) comprises 946 randomly generated simulation samples drawn uniformly within the same operating bounds as the DOE but not included in Dataset 1. These samples represent previously unseen operating conditions within the design domain and are used to evaluate the interpolation accuracy and in-domain generalization performance of the surrogate models.

- Dataset 3 (out-of-domain verification dataset) includes 345 randomly generated simulation samples, each containing at least one variable that extends beyond the DOE bounds by at least 20%. This dataset is used to assess the extrapolation behavior and robustness of the surrogate models when applied to operating conditions beyond those observed during training, thereby providing insight into model stability and potential limitations.
- Dataset 4 (optimization-focused dataset) consists of 150 simulation samples located in regions identified as optimal or near-optimal through Bayesian Optimization, covering multi-objective optimal operating zones. This dataset is used to verify the predictive accuracy of surrogate models and the reliability of the optimization framework in regions that are most relevant for process intensification and decision-making.

### 2.3.3. Data coverage and quality assessment

To quantitatively assess the quality and space-filling properties of the reduced DOE, a data coverage analysis was conducted using the star discrepancy metric (Niederreiter, 1992). Star discrepancy is a commonly used measure for evaluating the uniformity of point distributions in high-dimensional spaces and provides a quantitative indication of how well the sampled points represent the underlying design domain.

The coverage (discrepancy) is evaluated using the star discrepancy metric, formulated as a multivariate Kolmogorov–Smirnov–type statistic (Justel et al., 1997), as presented in Equation (1). The calculated star discrepancy indicates that the selected DOE samples achieve a coverage of approximately 52% relative to a hypothetical full-factorial design (for dataset 1) over the same operating ranges in this study. This result demonstrates that the reduced DOE provides a sufficiently uniform and representative sampling of the operating space despite the significant reduction in the number of simulation runs.

$$D_N^*(P) = \sup_{t \in [0,1]^d} \left| \frac{\#\{x_i \in [0,t)\}}{N} - \prod_{j=1}^{d} t_j \right| \quad (eq\ 1)$$

In addition to coverage assessment, basic data quality checks were performed to ensure consistency and reliability of the simulation results. These checks include verification of mass balance closure, physical feasibility of operating conditions for MeAce production, and consistency of performance indicators across the dataset. The combination of structured DOE, rigorous filtering of simulation results, and quantitative coverage assessment ensures that the generated dataset is suitable for robust surrogate model development and subsequent optimization studies.

## 2.4. Data-driven surrogate model development

### 2.4.1. Problem formulation

Based on the simulation dataset 1 generated through the DOE campaign, data-driven surrogate models were developed to approximate the steady-state behavior of the DME carbonylation process. The surrogate modeling task is formulated as a supervised regression problem, in which the selected operating variables serve as model inputs and key process performance indicators are treated as outputs.

Two separate regression models are constructed to capture different aspects of process performance:

- A surrogate model for predicting the C-yield, representing the MeAce production objective.
- A surrogate model for predicting the C-SEC, representing the overall energy efficiency of the process.

The input feature vector consists of the five operating variables (see Table 1) defined in Section 2.2. By decoupling the prediction of production and energy performance into two dedicated surrogate models, the framework enables flexible formulation of multi-objective optimization problems in later stages.

A total of 12 machine learning models were employed to explore different learning strategies. These models can be categorized as follows:

- Traditional models: Linear Regression (Hastie et al., 2009), Decision Tree (Breiman, 1998), Random Forest (Breiman, 2001). These models serve as baseline methods for comparison.
- Gradient boosting models: XGBoost (Chen and Guestrin, 2016), LightGBM (Ke et al., n.d.), CatBoost (Prokhorenkova et al., 2019), NGBoost (Duan et al., 2020). These ensemble methods are designed to improve prediction accuracy by sequentially combining weak learners.
- Tabular neural network models (Goodfellow et al., 2016; Rumelhart et al., 1986). were employed in this study to capture the highly nonlinear relationships between operating variables and process responses. A range of feed-forward neural network architectures with increasing structural complexity was investigated, spanning from shallow multilayer perceptrons (Popescu et al., n.d.) to deeper residual-based networks (He et al., 2015). In total, five neural network configurations—namely, MediumDNN, DeepNet, MLPEnsemble, Residual MLP, and DualHeadResNet—were trained and evaluated to assess the trade-offs between model complexity, predictive accuracy, and robustness. Detailed descriptions of the individual architectures and training procedures are provided in Appendix A. Among the evaluated models, a high-performing architecture referred to as DualHeadResNet was identified and selected for subsequent analysis and optimization. The overall structure of DualHeadResNet is shown in Figure 2. DualHeadResNet is based on a residual multilayer perceptron architecture (He et al., 2015; Popescu et al., n.d.)

comprising $N$ stacked residual blocks with layer normalization (Ba et al., 2016), linear transformations (Bishop, 2006), GELU activation functions (Hendrycks and Gimpel, 2023), and dropout regularization (Srivastava et al., n.d.). The use of residual connections facilitates gradient propagation in deep networks. DualHeadResNet produces its output as the sum of two components: a base head and a tail head. Both components are computed from a shared latent representation. The base head generates a scalar output that is passed through a sigmoid activation and scaled by a learnable parameter, resulting in a bounded baseline term. The tail head produces an additional scalar output, which is transformed using a softplus function to ensure non-negativity. The contribution of the tail component is controlled by a learnable scalar parameter passed through a sigmoid function, acting as a global gating factor. The final prediction is computed as the sum of the scaled base output and the gated tail component. This formulation represents the target as a combination of a bounded baseline term and a non-negative additive adjustment.

All models were configured with hyperparameters optimized for regression performance.

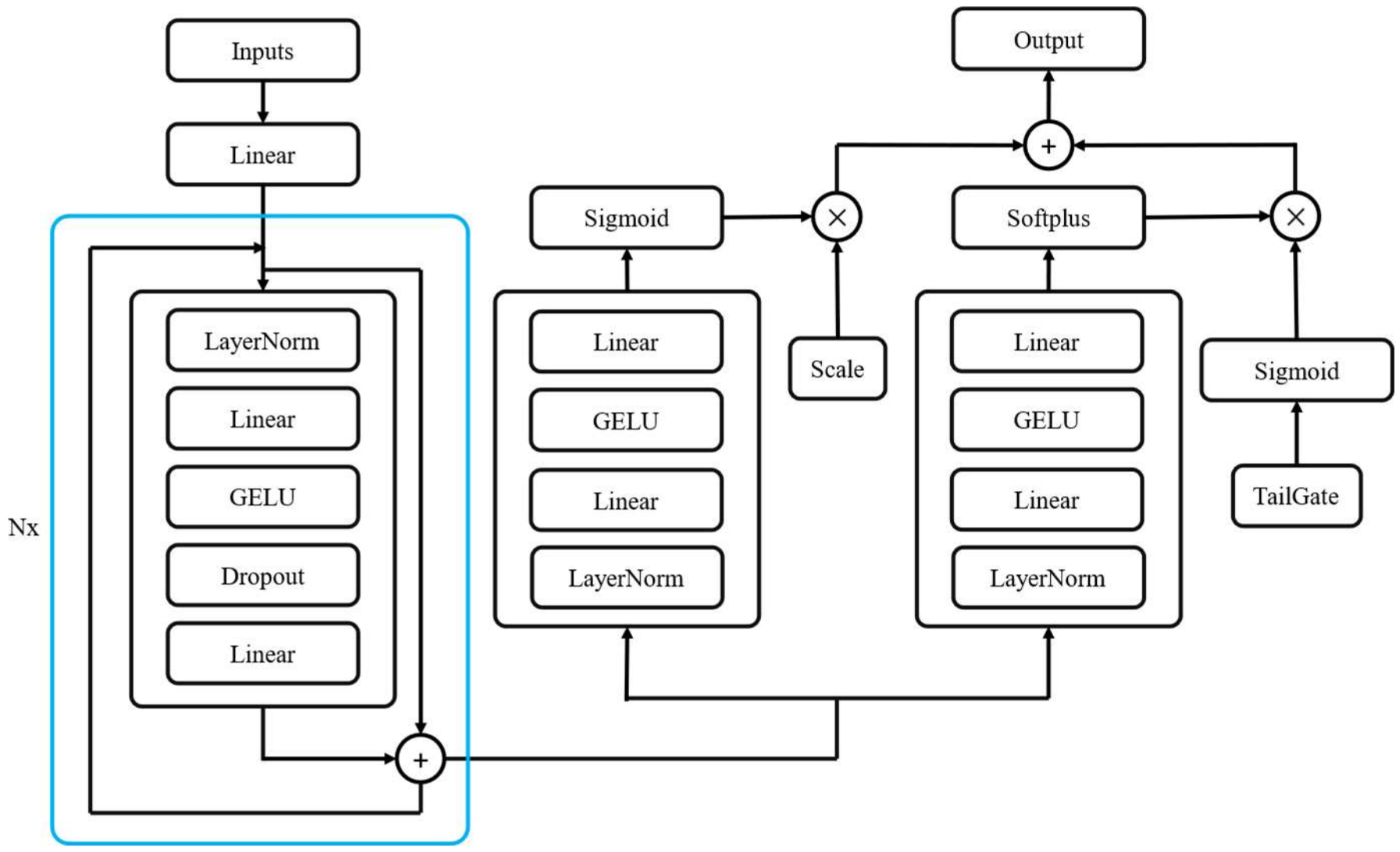

Figure 2: Structure of DualHeadResNet Model for C-yield and C-SEC

### 2.4.2. Model training and validation protocol

To ensure reliable predictive performance across the operating space, a structured training and validation protocol was employed. The primary simulation dataset (dataset1) was randomly divided into a training set comprising 85% of the data and a validation set comprising the remaining 15%. This split provides a balance between model training accuracy and independent performance assessment.

Beyond conventional random validation, additional verification datasets were generated to evaluate model robustness and generalization capability. These datasets include in-domain samples located within the original DOE bounds and out-of-domain samples representing operating conditions outside the training domain. This validation strategy allows systematic assessment of interpolation and extrapolation behavior, which is critical for surrogate models intended for optimization and decision-support applications.

Model performance was evaluated using multiple regression metrics, including the coefficient of determination ($R^2$), root mean absolute percentage error (MAPE), and mean absolute error (MAE). These metrics provide complementary insights into prediction accuracy, error magnitude, and robustness across different operating regimes. Emphasis was placed on consistent performance across validation datasets rather than solely maximizing accuracy on the training data.

### 2.4.3. Model selection for optimization

A range of machine learning model classes was evaluated to identify surrogate models suitable for optimization tasks. The evaluated models include traditional regression methods, ensemble-based models, and neural-network-based architectures designed for tabular data. The objective of this comparative evaluation was not to benchmark machine learning algorithms, but to identify surrogate models that offer an appropriate balance between prediction accuracy, numerical stability, and generalization capability.

Model selection was guided by three main criteria: predictive accuracy on validation datasets, robustness under in-domain and out-of-domain conditions, and smoothness of the predicted response surface. The latter is particularly important for optimization, as unstable or highly irregular surrogate predictions can lead to unreliable optimization outcomes.

Based on these criteria, a representative surrogate model was selected for integration into the Bayesian optimization framework. Detailed comparisons of individual model performances, including additional evaluation metrics and parity plots, are provided in the Appendix A. The selected surrogate models serve as computationally efficient approximations of the underlying process simulation and form the basis for subsequent process intensification and optimization studies.

## 2.5. Bayesian optimization framework

### 2.5.1. Optimization problem formulation

The process intensification problem is formulated as an optimization task in the space of operating variables defined in Section 2.2, including the reactor length, the reactor temperature, the system pressure, the CO molar flow rate, and the splitter ratio. All variables are bounded within physically feasible and industrially relevant ranges to ensure meaningful and realistic optimization outcomes.

A multi-objective formulation is adopted to simultaneously maximize carbon yield (C-yield) and minimize the cost of specific energy consumption (C-SEC). While these objectives exhibit a favorable coupling at moderate conversion levels, a clear trade-off emerges near the Pareto-optimal region, where further gains in C-yield require disproportionately higher energy input. This behavior reflects the intrinsic interaction between reaction conversion, recycle flow rates, and system-level energy demand.

Process feasibility and numerical stability are implicitly enforced through the surrogate models, which are trained exclusively on physically feasible and converged simulation data. As a result, the optimization is naturally constrained to realistic and operable regions of the design space without the need for explicit constraint handling.

### 2.5.2. Surrogate-based optimization strategy

Bayesian Optimization (BO) is employed as the optimization strategy due to its suitability for high-dimensional problems with expensive function evaluations. In the present framework, the surrogate models developed in Section 2.4 replace direct process simulations as evaluators of the objective functions. This surrogate-based approach significantly reduces computational cost while enabling efficient exploration of the operating space.

The BO framework follows an iterative procedure in which new candidate operating points are proposed based on a balance between exploration of uncertain regions and exploitation of promising regions identified by the surrogate models (Chatterjee et al., 2019; Paria et al., 2019a; Zhou et al., 2024). At each iteration, the surrogate predictions are used to estimate objective values and associated uncertainty, guiding the search toward optimal regions of the design space.

For multi-objective optimization, BO is extended to handle multiple conflicting objectives by identifying a set of Pareto-optimal solutions rather than a single optimal point. This approach enables systematic analysis of trade-offs between C-yield and C-SEC, providing insight into the structure of the Pareto front and the sensitivity of optimal solutions to operating conditions (Knowles, 2006).

To facilitate scalarization-based optimization, the multi-objective problem is also reformulated into a single composite objective using a linear scalarization approach (Chugh, 2019; Paria et al., 2019b), following common practice in multi-objective Bayesian optimization

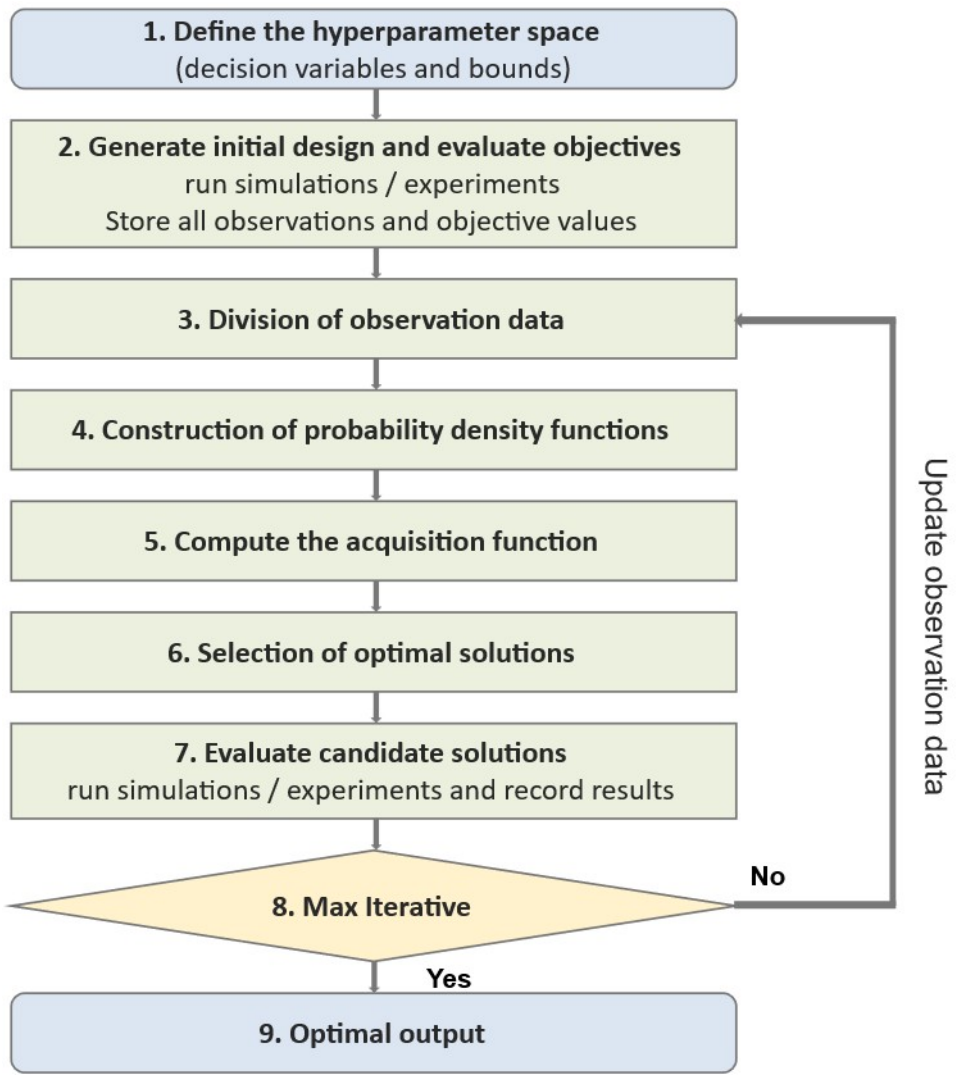

Figure 3: Bayesian Optimization framework using the TPE algorithm adopted from a literature (Zhang et al., 2025)

The optimization strategy focuses on aggregating the two objectives: C-yield and C-SEC into a single target function to be optimized. Prior to aggregation, C-yield and C-SEC are normalized using the minimum and maximum values obtained from the DOE dataset to ensure comparable scaling of objectives. The normalized objectives are not restricted to the [0,1] interval during optimization, allowing candidate solutions to exceed the initial DOE bounds and enabling the identification of improved operating conditions. While scalarization is used to guide single-objective optimization, Pareto-optimal solutions are retained to analyze trade-offs between competing objectives, consistent with established multi-objective optimization frameworks (Knowles, 2006).

Specifically, the composite objective is defined as an improvement score (equation 2), expressed as

$$Improvement\ Score = w_{C_yield} . C_yield_{scaled} + w_{C-S} \quad . (1 - C_SEC_{scaled}) \text{ (eq 2)}$$

where $C_yield_{scaled}$ and $C_SEC_{scaled}$ are obtained via min–max normalization using the minimum and maximum values from the DOE dataset. $w_{C_yield}$ , $w_{C-SEC}$ are adopted to balance the objectives of maximizing C-yield and minimizing C-SEC.

With this formulation, the improvement score increases only when both objectives move toward their respective optimal directions, promoting balanced improvements rather than favoring one objective at the expense of the other. As a result, improvements in the objective function are achieved only when operating conditions simultaneously enhance C-yield and reduce C-SEC, or when an optimal trade-off between the two is established. This formulation guides the optimization toward balanced solutions and reflects the intrinsic trade-off structure of process intensification.

The BO procedure is implemented using the Optuna framework (Akiba et al., 2019), which provides an efficient and flexible platform for multi-objective optimization. The Tree-structured Parzen Estimator (TPE) (Bergstra et al., n.d.; Watanabe, 2025) sampler is adopted to guide the optimization process by probabilistically modeling promising and less promising regions of the operating space, while avoiding unnecessary evaluations in low-performing regions of the design space.

### 2.5.3. Optimization scenarios

The workflow of the Bayesian optimization procedure is illustrated in Figure 3. BO is an iterative optimization approach that sequentially searches for optimal solutions based on previously evaluated operating points (Bergstra et al., n.d.; Watanabe, 2025; Zhang et al., 2025). At each iteration, the method proposes new operating conditions by balancing the improvement of objective functions with exploration of the operating space. The optimization process progressively refines the search toward high-performing regions as more information becomes available. This strategy allows efficient convergence to optimal solutions with a limited number of evaluations.

The multi-objective optimization scenario addresses the simultaneous optimization of C-yield and C-SEC. Instead of yielding a single optimal solution, this scenario generates a set of Pareto-optimal operating points that represent different trade-offs between productivity and energy efficiency. These solutions allow identification of operating regimes that offer balanced performance improvements and support informed decision-making based on specific operational priorities. The optimization results obtained from this study for subsequent process intensification analysis and are further interpreted in the results and discussion section.

## 2.6. Scope of the methodology

The methodology presented in this study is subject to several scope definitions and limitations that should be clearly acknowledged when interpreting the results. These limitations are inherent to the objectives and methodological choices of the work and do not detract from its intended contribution to process intensification studies.

The data used for surrogate model development and optimization are entirely derived from steady-state process simulations. No experimental or plant data are included in the present study. As a result, the quantitative predictions obtained from the surrogate models and optimization framework are constrained by the fidelity of the underlying process simulation model. The proposed methodology is therefore intended for operating-window exploration, screening, and decision support, rather than as a direct replacement for experimental validation.

The base process configuration, reaction scheme, and kinetic description are adopted from previously published literature (Diemer and Luyben, 2010). The reaction mechanism and kinetic models are not modified in this work. This choice is made deliberately to focus on operating-condition optimization and energy–performance trade-offs rather than on flowsheet synthesis, catalyst development, or kinetic modeling. Consequently, the novelty of the present study lies in the integration of large-scale simulation data, data-driven surrogate modeling, and optimization for process intensification, rather than in the development of new process chemistry or reaction models.

The process simulations are performed under steady-state assumptions. Dynamic behavior, transient operation, and control-related aspects are not considered. While dynamic effects may be important for start-up, shut-down, or disturbance handling, the steady-state framework adopted here is appropriate for analyzing nominal operating conditions and identifying optimal operating regions for intensification purposes.

Within these defined scope boundaries, the proposed methodology provides a systematic, scalable, and computationally efficient approach for integrating process simulation, data-driven modeling, and optimization. The framework is readily extendable to other reaction–separation systems and can be complemented in future work by experimental validation, dynamic modeling, or alternative process configurations.

# 3 Results and discussion

### 3.1. Process behavior and operating-window insights from DOE

The large-scale DOE-based steady-state simulation dataset was first analyzed to elucidate the underlying process behavior and to identify key operating windows governing C-yield and C-SEC. The analysis presented in this section is based on Dataset 1, comprising 2730 converged simulation cases, which provides a space-filling representation of the five-dimensional operating space.

Figure 4 illustrates the combined effects of reactor temperature, reactor length, CO molar flow rate, and splitter ratio on C-yield. Among all variables, reactor temperature emerges as the dominant factor governing the overall conversion behavior.

At low reactor temperatures (below approximately 451 K), the C-yield remains uniformly low across the entire operating space, with maximum values not exceeding ~39%. This behavior indicates a kinetically limited regime, where insufficient thermal activation suppresses the carbonylation pathway regardless of variations in CO availability or residence time.

As the reactor temperature increases beyond 451 K, a pronounced improvement in C-yield is observed. In particular, within the temperature range of approximately 477–486 K, favorable combinations of operating conditions enable C-yield values to reach up to 85–88%. This region represents a high-performance operating window in which reaction kinetics are sufficiently activated to promote efficient carbon conversion.

However, it is important to note that increasing temperature may also accelerate coke formation and catalyst deactivation (Cheng et al., 2017), potentially limiting long-term process performance. Although catalyst deactivation and coke formation are not explicitly considered in the present steady-state framework, these effects are expected to impose an upper bound on the practically achievable temperature range in industrial applications.

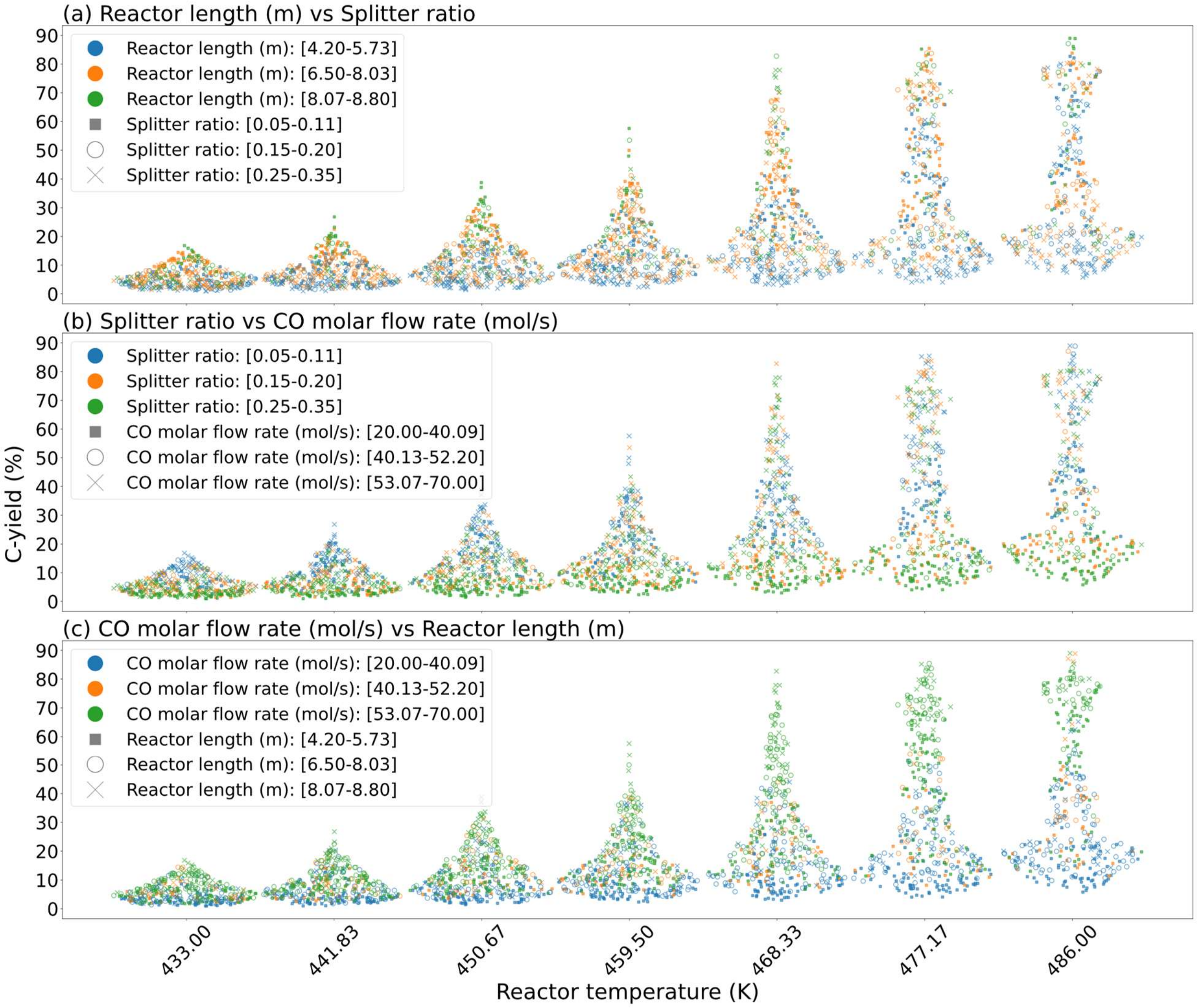

**Figure 4.** *Effects of reactor temperature, reactor length, splitter ratio, and CO molar flow rate on C-yield.*

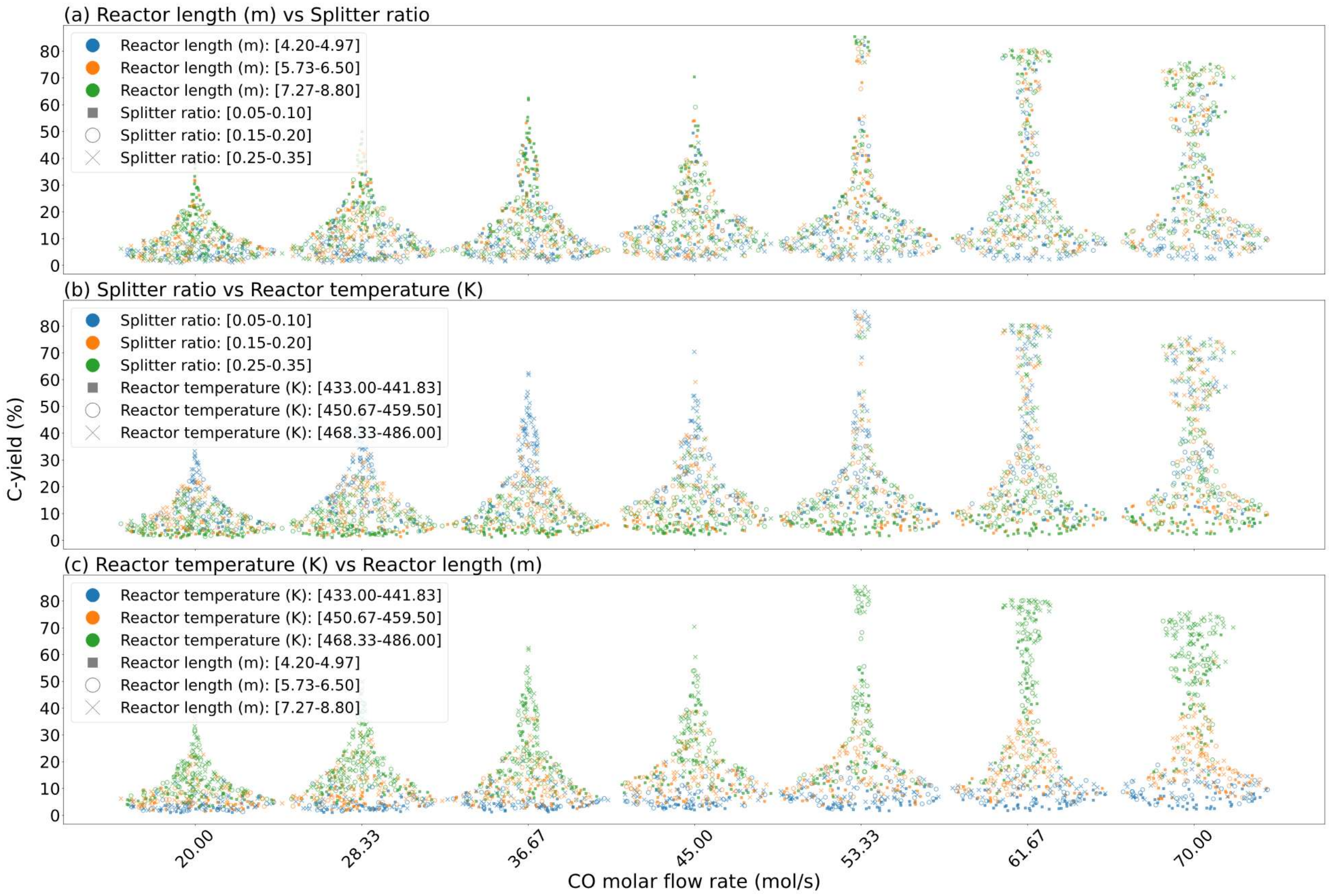

**Figure 5.** *Effects of CO molar flow rate, reactor length, splitter ratio, and CO molar flow rate on C-yield.*

The influence of CO molar flow rate on C-yield is further illustrated in Figure 5. An increase in CO flow rate generally leads to a corresponding increase in C-yield, reflecting enhanced availability of carbonylation reactants. However, this trend is not monotonic across the entire range. The C-yield reaches a maximum (approximately 85–88%) at CO flow rates around 53–54 mol·s$^{-1}$ under favorable conditions. Beyond this range, further increases in CO flow result in a slight decrease in C-yield (on the order of ~10%), indicating diminishing returns and potential inefficiencies due to excess reactant. This behavior suggests the existence of an optimal CO operating window centered around 53–54 mol·s$^{-1}$ for the fixed DME feed conditions considered in this study. Reactor length also exhibits a clear and systematic influence on C-yield, as observed in Figures 4c and 5c. C-yield increases noticeably when the reactor length exceeds approximately 5.7 m. This trend can be attributed to the increase in residence time, which allows more complete utilization of reactants and promotes higher overall conversion. Longer reactors therefore contribute to improved process performance by enabling the reaction system to approach its intrinsic kinetic potential. In contrast, the effect of the splitter ratio of the purge gas is not clearly pronounced within the investigated range (5–35%). Across the 2730 simulation cases, no consistent monotonic trend is observed between splitter ratio and C-yield. This suggests that, within the studied operating domain, the splitter ratio is strongly coupled with other variables, rather than acting as a primary driver of conversion. Based on the identification of high-performance regions with C-yield values in the range of 83–88%, the favorable operating window for C-yield can be approximately defined as follows: CO molar flow rate around 53 $mol \cdot s^{-1}$, reactor temperature in the range of 468–486 K, and reactor length between 5.7 and 8.8 m. These conditions represent a region where kinetic activation, reactant availability, and residence time are jointly optimized.

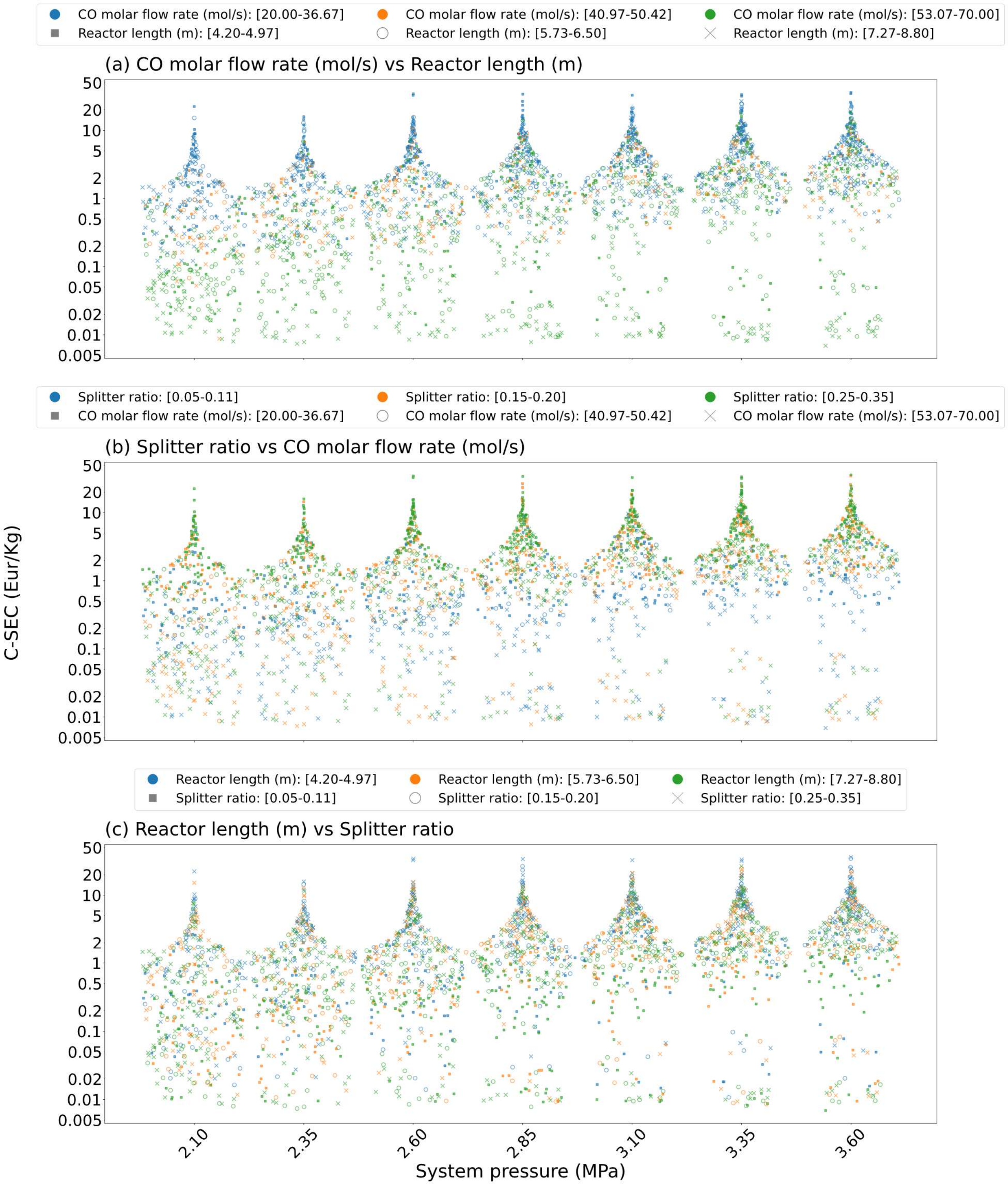

**Figure 6.** *Effects of system pressure, CO molar flow rate, reactor length and splitter ratio on C-SEC.*

The effects of operating variables on C-SEC are presented in Figures 6 and 7. System pressure has a negligible impact on C-SEC within the investigated range of 2.1–3.6 MPa. This observation is consistent with the underlying process simulation, where the energy consumption associated with feed and recycle gas compression is significantly lower than the steam demand required for the distillation column (the data are not presented in this study). As a result, variations in system pressure do not substantially affect the overall energy cost (see Figure 6). Similar to its effect on C-yield, CO molar flow rate plays a critical role in determining C-SEC. To achieve energy consumption levels below approximately 0.1 $EUR \cdot kg^{-1}$ MeAce, the CO flow rate must exceed ~53 $mol \cdot s^{-1}$ (see Figures 6a and 7a). This behavior is directly linked to improved conversion efficiency at optimal CO levels, which reduces recycling flow rates and increases MeAce production, thereby lowering the cost of energy consumption per unit of product. Reactor temperature also exhibits a strong influence on C-SEC, following a trend similar to that observed for C-yield. Increasing temperature leads to a significant reduction in C-SEC, primarily due to enhanced conversion of reactants into MeAce. Higher conversion lowers the overall energy demand associated with recycle compression and separation.

The DOE-based analysis provides several important insights relevant to process intensification. The operating space contains distinct high-performance regions characterized by simultaneous improvement in C-yield and C-SEC. These regions occupy only a small fraction of the overall design space, highlighting the limitations of conventional trial-and-error optimization approaches. The strong interaction effects observed among temperature, reactor length, CO flow rate, and splitter ratio indicate that multivariate optimization strategies are essential for effective operating-window identification. This observation highlights a key characteristic of the DME carbonylation process: energy efficiency is largely governed by conversion performance. Consequently, process intensification strategies aimed at maximizing MeAce production inherently lead to improved energy performance, provided that operating conditions remain within feasible and stable limits.

In summary, the optimization of both C-yield and C-SEC in this study largely depends on maximizing MeAce production. Higher MeAce production directly reflects improved C-yield, which reduces the amount of unreacted species in the recycle loop and consequently lowers the overall energy demand associated with compression and separation. As a result, operating conditions that favor high conversion not only enhance process performance but also inherently improve energy efficiency, highlighting a strong coupling between production intensification and energy consumption

### 3.2. Performance of data-driven surrogate models

The performance of data-driven surrogate models, trained exclusively on process simulation data and employed as computationally efficient proxies for process models, was evaluated systematically to assess their suitability for process optimization and decision-support applications. A total of 12 machine learning models were investigated for predicting both C-yield and C-SEC. Model performance was assessed using three complementary datasets: the validation subset of the DOE dataset (Dataset 1 with 2730 simulations), the in-domain verification dataset (Dataset 2), and the out-of-domain verification dataset (dataset 3).

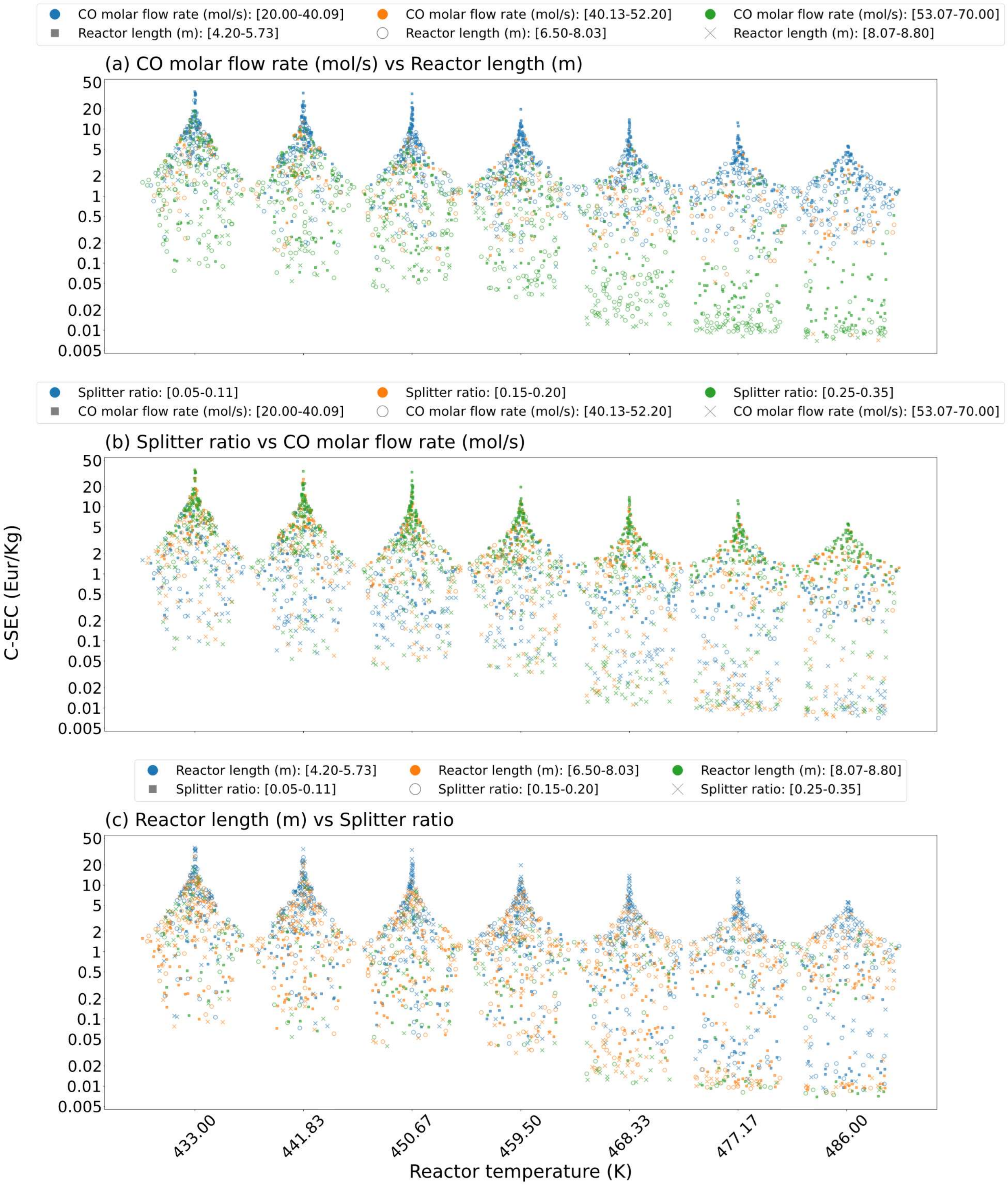

**Figure 7.** *Effects of system pressure, CO molar flow rate, reactor length and splitter ratio on C-SEC.*

Model performance was quantified using the coefficient of determination ($R^2$), mean absolute error (MAE), and mean absolute percentage error (MAPE). $R^2$ evaluates the goodness-of-fit and variance explained by each model, while MAE provides an interpretable measure of absolute prediction error in physical units. MAPE complements these metrics by assessing relative prediction accuracy in percentage terms, enabling robust comparison across datasets and operating regimes. The predictive performance of surrogate models for C-yield was evaluated using three complementary datasets: the validation subset of the DOE dataset, an in-domain verification dataset, and an out-of-domain verification dataset. In total, 12 machine learning models were assessed, covering a broad spectrum of learning strategies, including traditional regression models, gradient boosting methods, and tabular deep learning architectures. A comparative summary of model performance across these three model families is provided in Table 2.

**Table 2.** Comparison on Validation in-bound and out-bound set of C-yield model

| Family | Model | Validation Set | | | In-Bound Set | | | Out-Bound Set | | |
|---|---|---|---|---|---|---|---|---|---|---|
| | | MAE | R2 | MAPE | MAE | R2 | MAPE | MAE | R2 | MAPE |
| Neural Network | DualHeadResNet | 3.8e-3 | 0.97 | 1.48 | 3.1e-3 | 0.97 | 1.33 | 1.0e-2 | 0.97 | 6.98 |
| | Residual MLP | 6.1e-3 | 0.97 | 4.31 | 6.0e-3 | 0.97 | 3.72 | 1.3e-2 | 0.90 | 14.87 |
| | MLP Ensemble | 1.0e-2 | 0.97 | 8.98 | 1.3e-2 | 0.95 | 7.92 | 2.0e-2 | 0.95 | 17.82 |
| | DeepNet | 1.1e-2 | 0.96 | 9.90 | 2.0e-2 | 0.92 | 7.93 | 2.7e-2 | 0.92 | 18.09 |
| | Medium DNN | 1.5e-2 | 0.96 | 15.56 | 2.6e-2 | 0.90 | 12.26 | 3.5e-2 | 0.91 | 24.88 |
| Boosting-based | CatBoost | 7.4e-3 | 0.97 | 3.89 | 2.0e-2 | 0.93 | 12.23 | 5.0e-2 | 0.86 | 44.70 |
| | LightGBM | 1.2e-2 | 0.95 | 7.80 | 3.0e-2 | 0.81 | 17.75 | 5.3e-2 | 0.84 | 44.82 |
| | XGBoost | 1.2e-2 | 0.96 | 8.01 | 3.6e-2 | 0.84 | 20.40 | 5.8e-2 | 0.82 | 45.03 |
| | NGBoost | 1.9e-2 | 0.94 | 13.04 | 3.3e-2 | 0.88 | 20.20 | 5.9e-2 | 0.83 | 49.04 |
| Conventional ML | Random Forest | 1.5e-2 | 0.95 | 9.08 | 2.8e-2 | 0.87 | 16.50 | 5.8e-2 | 0.84 | 52.24 |
| | Decision Tree | 2.1e-2 | 0.91 | 13.61 | 3.5e-2 | 0.83 | 20.43 | 5.8e-2 | 0.81 | 49.99 |
| | Linear Regression | 5.8e-2 | 0.76 | 63.74 | 5.8e-2 | 0.68 | 44.15 | 8.6e-2 | 0.69 | 114.99 |

Clear differences in predictive accuracy and robustness are observed among the evaluated model families, underscoring the importance of nonlinear representation capacity for capturing the complex process behavior governing C-yield. Traditional regression models, such as linear regression and single decision trees, exhibit limited predictive accuracy, particularly under in-domain and out-of-domain conditions, as indicated by relatively low $R^2$ values and high error MAE and MAPE metrics. This behavior reflects their inability to adequately represent the strong nonlinear interactions among operating variables. Gradient boosting models demonstrate substantially improved performance on the validation dataset, with CatBoost achieving an $R^2$ value of 0.97 and relatively low MAE and MAPE values. However, their predictive performance deteriorates noticeably when applied to operating conditions outside the training domain, as evidenced by increased prediction errors and reduced $R^2$ values on the out-of-domain dataset. This degradation highlights limitations in extrapolation robustness for tree-based ensemble methods in highly nonlinear process systems. In contrast, deep learning–based models exhibit superior and more consistent performance across all datasets. These models leverage multi-layer neural network architectures to learn complex nonlinear mappings between operating variables and process responses. Among the evaluated architectures, DualHeadResNet achieves the highest predictive accuracy, with an $R^2$ of 0.97 and an MAPE of 1.48 on the validation dataset, while maintaining robust performance under both in-domain and out-of-domain conditions.

The validation results for both in-domain and out-of-domain datasets are illustrated in Figure 8, which compares the predictions of four representative machine learning models against the CoCo process simulation under identical operating conditions. The color scale in the figure represents the prediction error between the surrogate models and the process simulation results. A complete comparison of all 12 evaluated models is provided in Appendix A.

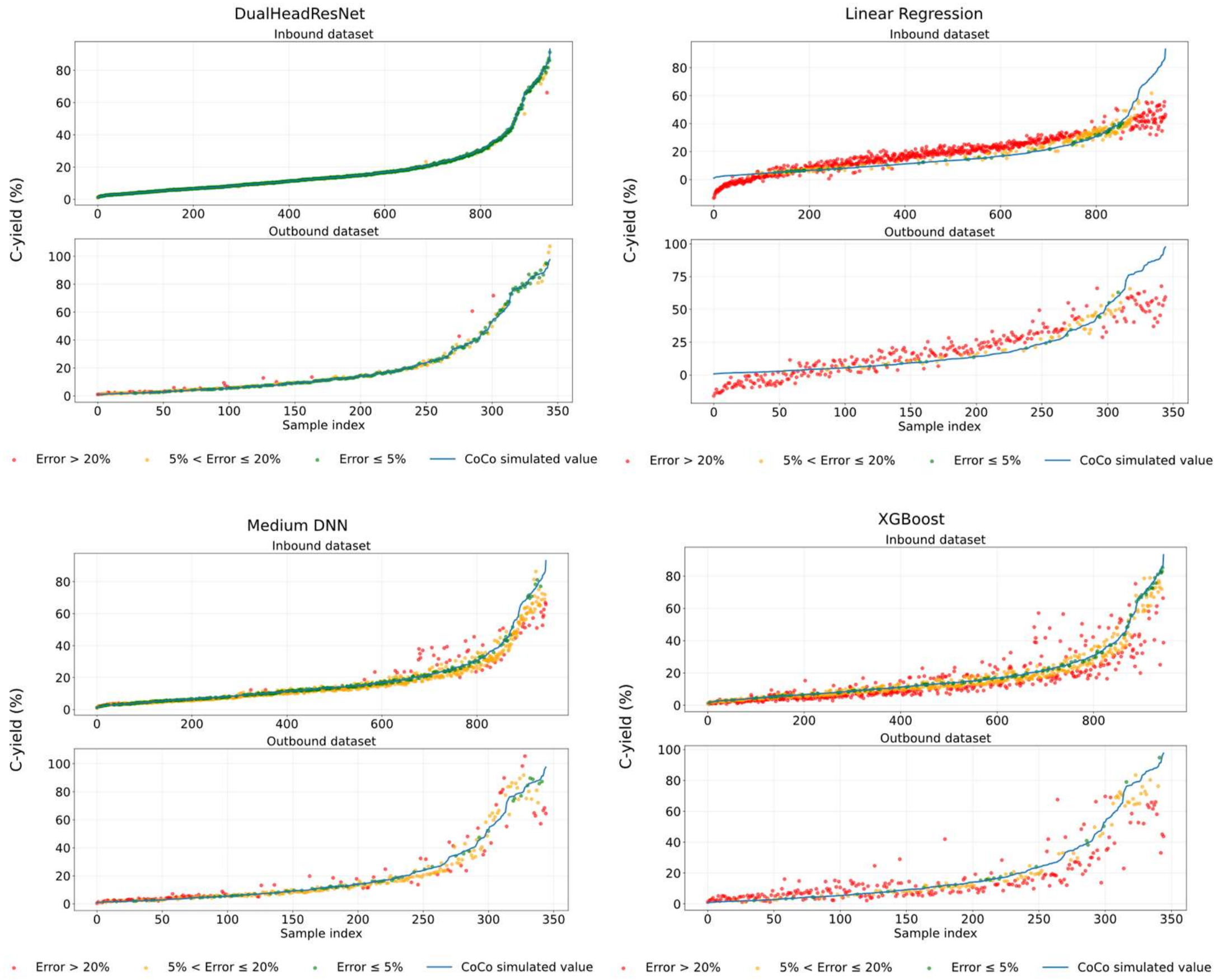

**Figure 8.** Validation of surrogate models on in-domain and out-of-domain datasets for C-yield. ERROR denotes the MAPE metrics between CoCo process simulation and machine learning models.

A quantitative assessment of model performance was conducted using 946 in-domain samples and 345 out-of-domain samples. Across both datasets, a consistent error distribution pattern is observed for all model families, indicating that the relative performance differences among models are preserved when transitioning from interpolation to extrapolation conditions. Although prediction errors increase under out-of-domain conditions, the overall structure of the error distribution remains similar, suggesting that model behavior is systematic rather than random.

Among the evaluated models, neural network–based architectures with higher structural complexity demonstrate superior predictive performance. These models are able to capture the highly nonlinear relationships governing C-yield, resulting in consistently lower prediction errors across the entire response range—from low to high C-yield—when compared to CoCo simulation results. In contrast, traditional regression models and several tree-based approaches exhibit larger deviations and fail to accurately reproduce nonlinear trends, particularly near the boundaries of the operating domain.

**Table 3.** Comparison on Validation in-bound and out-bound set of C-SEC model

| Family | Model | Validation Set | | | In-Bound Set | | | Out-Bound Set | | |
|---|---|---|---|---|---|---|---|---|---|---|
| | | MAE | R2 | MAPE | MAE | R2 | MAPE | MAE | R2 | MAPE |
| Neural Network | DualHeadResNet | 5.5e-2 | 0.97 | 2.18 | 5.4e-2 | 0.97 | 2.35 | 3.9e-1 | 0.94 | 11.28 |
| | Residual MLP | 8.3e-2 | 0.97 | 13.17 | 7.2e-2 | 0.97 | 7.62 | 6.1e-1 | 0.90 | 23.87 |
| | Medium DNN | 1.4e-1 | 0.97 | 23.67 | 4.4e-1 | 0.83 | 14.80 | 1.4e+0 | 0.68 | 29.54 |
| | DeepNet | 3.4e-1 | 0.93 | 17.81 | 2.5e-1 | 0.94 | 19.79 | 9.9e-1 | 0.82 | 30.90 |
| | MLP Ensemble | 2.5e-1 | 0.95 | 51.96 | 3.4e-1 | 0.88 | 24.01 | 1.3e+0 | 0.71 | 76.05 |
| Boosting-based | CatBoost | 9.8e-2 | 0.96 | 20.02 | 3.7e-1 | 0.92 | 41.77 | 1.9e+0 | 0.54 | 194.60 |
| | LightGBM | 1.3e-1 | 0.97 | 31.34 | 5.2e-1 | 0.86 | 57.02 | 1.9e+0 | 0.56 | 204.34 |
| | NGBoost | 1.6e-1 | 0.97 | 53.56 | 4.5e-1 | 0.87 | 61.73 | 2.0e+0 | 0.54 | 256.68 |
| | XGBoost | 1.5e-1 | 0.97 | 38.25 | 9.8e-1 | 0.46 | 101.66 | 1.9e+0 | 0.62 | 255.58 |
| Conventional ML | Random Forest | 4.5e-1 | 0.90 | 31.79 | 4.7e-1 | 0.89 | 44.04 | 2.2e+0 | 0.46 | 267.95 |
| | Decision Tree | 7.1e-1 | 0.85 | 37.10 | 7.6e-1 | 0.75 | 43.81 | 2.3e+0 | 0.44 | 287.35 |
| | Linear Regression | 8.1e-1 | 0.74 | 257.71 | 5.9e-1 | 0.75 | 201.54 | 1.8e+0 | 0.56 | 591.44 |

These results highlight the advantage of advanced neural network architectures in representing complex process behavior. In particular, DualHeadResNet achieves the highest level of extrapolation robustness, with an $R^2$ of approximately 0.97 and consistently lower error metrics across both datasets. This performance demonstrates its ability to generalize beyond the DOE domain and confirms its suitability as a reliable surrogate model for subsequent optimization and decision-support applications. Unlike experimental data, typical sources of noise are absent in this steady-state simulation environment. The training data were generated from a DOE-based process simulation framework with approximately 52% coverage of the design space. As a result, the high predictive accuracy of the models facilitates a clearer and more systematic analysis of the interaction effects among the studied variables on both C-yield and C-SEC.

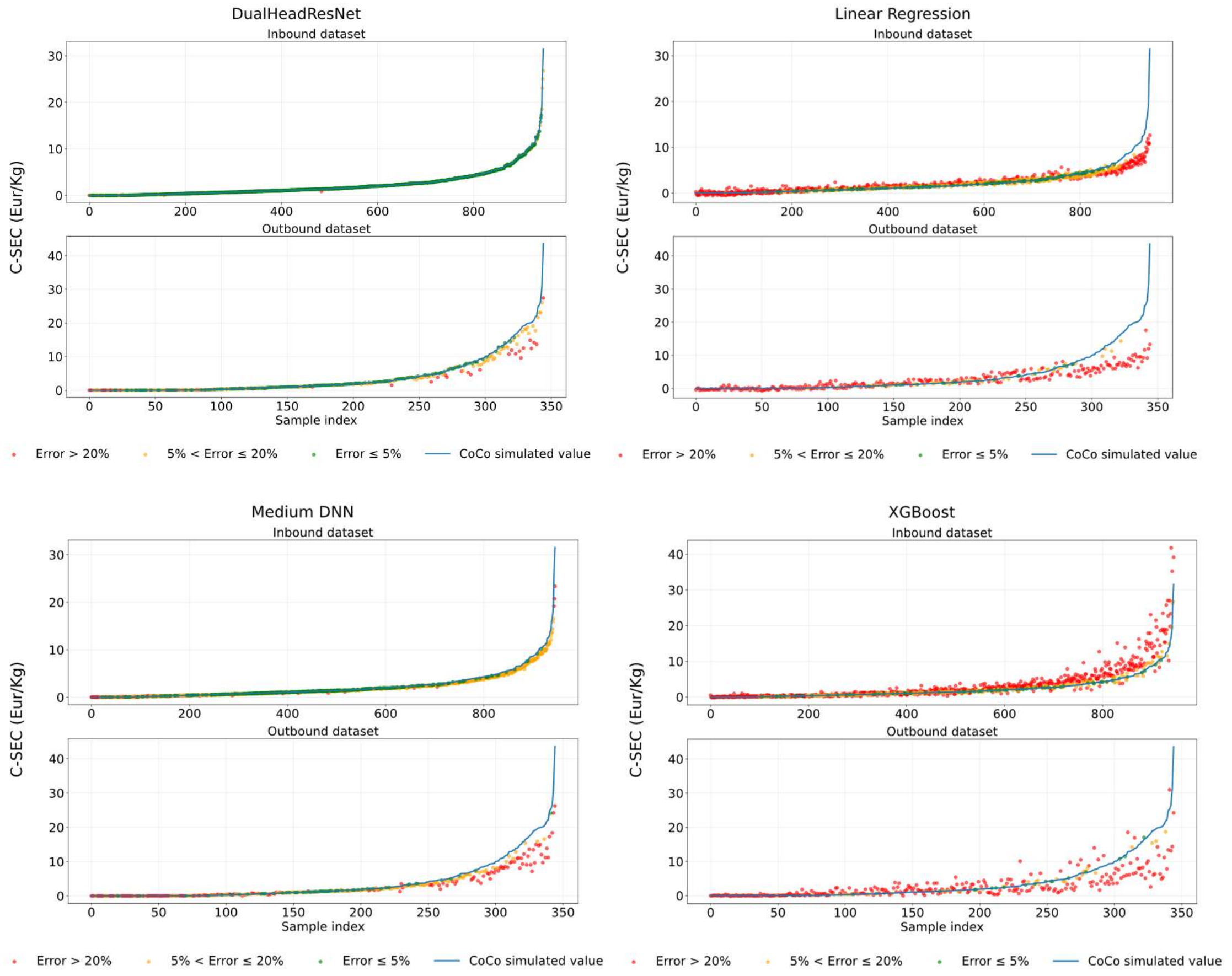

**Figure 9.** Validation of surrogate models on in-domain and out-of-domain datasets for C-SEC. ERROR denotes the MAPE metrics between CoCo process simulation and machine learning models.

This superior performance can be directly linked to the underlying model architecture. DualHeadResNet is designed to balance smooth global approximation with localized modeling of extreme responses. By decomposing the prediction into a bounded base component and a gated tail component, the model effectively captures the dominant trends of C-yield while selectively accounting for high-response regimes. In addition, the residual structure enhances training stability and facilitates generalization, while the adaptive gating mechanism regulates the contribution of the tail component to prevent overfitting and maintain physical consistency. As a result, DualHeadResNet delivers stable and accurate predictions across all evaluation datasets, with low sensitivity to distribution shifts between in-domain and out-of-domain conditions. This robustness reinforces the suitability of advanced deep learning–based surrogate models for representing nonlinear process behavior. More importantly, it provides a reliable foundation for surrogate-assisted Bayesian optimization frameworks, where consistent model behavior across the operating space is essential.

Table 3 summarizes the predictive performance of surrogate models for C-SEC across the validation, in-domain, and out-of-domain datasets. Compared to C-yield, C-SEC represents a more aggregated and system-level response, as it arises from the combined effects of cost from compression, recycling flow, heat integration and pressure-related losses across multiple unit operations. As a result, accurate prediction of C-SEC requires surrogate models capable of capturing complex nonlinear interactions at the process level.

Across all evaluation datasets, neural network–based models demonstrate superior performance relative to traditional regression and gradient boosting approaches, achieving high predictive accuracy on the validation dataset. Among the evaluated models, DualHeadResNet exhibits the most stable and consistent behavior, maintaining high accuracy under both in-domain ($R^2 = 0.97$) and out-of-domain ($R^2 = 0.94$) conditions with comparatively low prediction errors. In contrast, gradient boosting and conventional models show a marked deterioration in extrapolation performance, indicating limited robustness when applied beyond the training domain. These results highlight the advantage of advanced nonlinear surrogate models for representing system-level energy behavior and support their use in surrogate-based optimization and multi-objective analyses targeting energy efficiency.

Figure 9 presents a detailed comparison of surrogate model predictions for C-SEC against CoCo process simulation results under both in-domain and out-of-domain conditions. Similar to the C-yield evaluation, the figure illustrates the performance of four representative machine learning models under identical operating conditions, with the color scale indicating the prediction error between surrogate models and simulation results.

Consistent with the observations for C-yield, the evaluation of C-SEC across both in-domain and out-of-domain datasets reveals similar error distribution patterns among the different model families. Although prediction errors increase under extrapolation conditions, the overall structure of the error distribution remains comparable between the two regions, indicating stable and systematic model behavior. Among the evaluated models, neural network–based architectures clearly outperform gradient boosting and conventional regression models across the entire range of C-SEC values. In particular, these models maintain high predictive accuracy for both low and high C-SEC regimes, demonstrating their ability to capture the complex, system-level interactions governing energy consumption. In contrast, tree-based and traditional models exhibit larger error dispersion, especially at the extremes of the response range, reflecting their limited capability to generalize under highly nonlinear and extrapolative conditions. These visual observations are consistent with the quantitative metrics reported in Table 3 and further confirm the superior robustness and generalization capability of deep learning–based surrogate models. In particular, DualHeadResNet provides the most accurate and stable predictions across both datasets, reinforcing its suitability for surrogate-based optimization targeting energy-efficient process intensification.

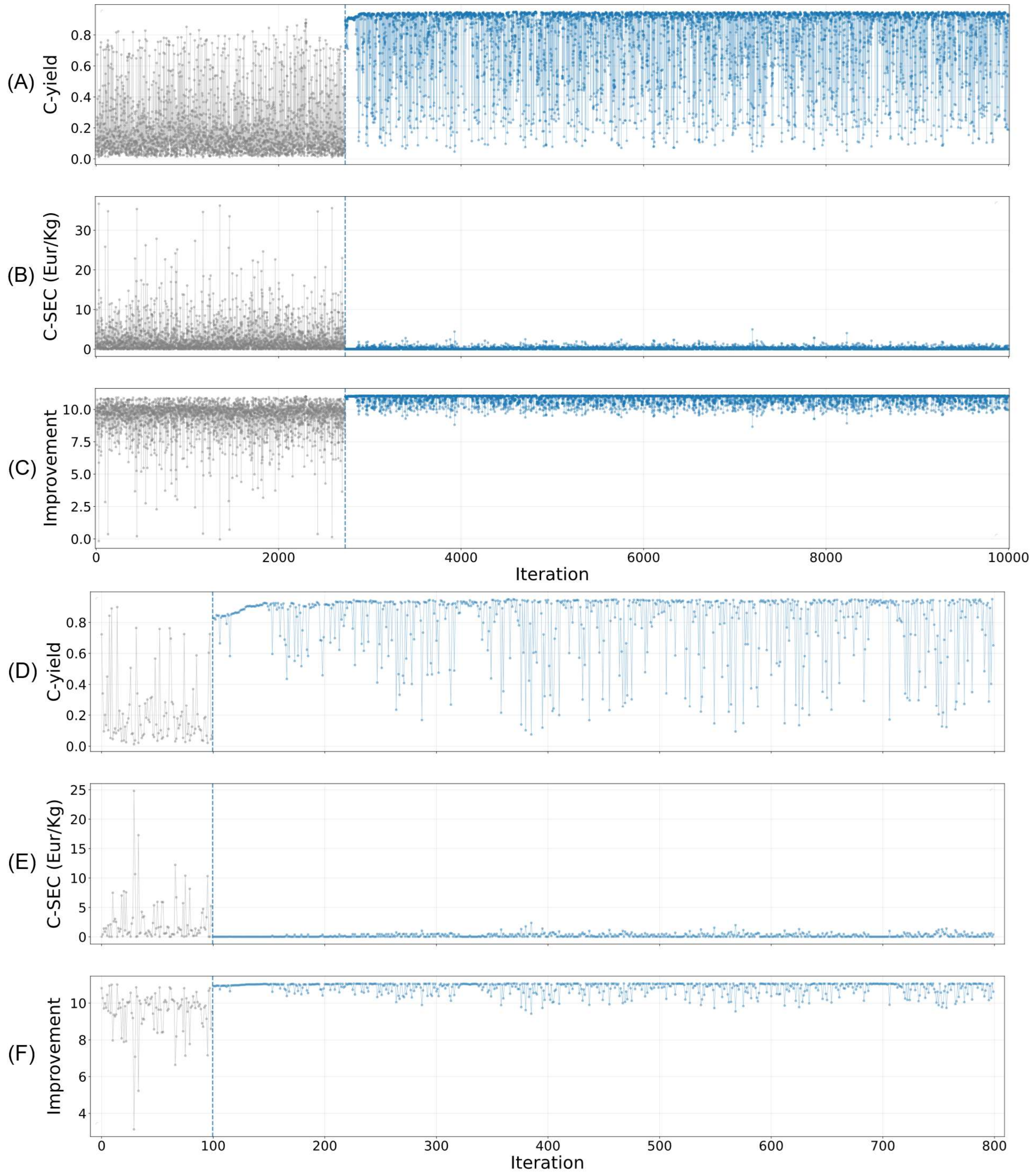

**Figure 10.** Convergence behavior of multi-objective Bayesian optimization for C-yield and C-SEC under two optimization scenarios - (A-C): Case 1 and (D-F): Case 2.

Based on the combined evaluation across validation, in-domain, and out-of-domain datasets, the DualHeadResNet model is selected as the primary surrogate for both C-yield and C-SEC. Although some models achieve slightly higher accuracy on the validation dataset, DualHeadResNet consistently demonstrates a favorable balance between predictive accuracy, stability, and robustness across all datasets. This balanced performance is particularly important for Bayesian optimization, where surrogate models are repeatedly queried in regions that may not be densely sampled during training. Consequently, DualHeadResNet is integrated into the optimization framework to predict C-yield and C-SEC under different operating conditions and to support multi-objective optimization analyses presented in the following sections.

### 3.3. Multi-objective optimization for methyl acetate production and cost of energy consumption

Multi-objective Bayesian optimization was employed to simultaneously maximize C-yield and minimize C-SEC. To assess the influence of data availability and optimization budget on the resulting Pareto front, two representative case studies, DOE-intensive optimization and data-efficient optimization, were investigated. As described in Section 2.5.2, the optimization is performed by transforming the two objectives into a single-scale objective, thereby reformulating the original multi-objective optimization problem into a single-objective maximization.

In DOE-intensive optimization case study (Case 1), the optimization was initialized using the full DOE dataset comprising 2730 simulation points, followed by 10,000 BO iterations. This configuration represents an extensive exploration strategy aimed at approximating a near-global Pareto front. In contrast, data-efficient optimization case study (Case 2) was initialized with only 100 randomly sampled points and performed 800 BO iterations, reflecting a more practical optimization scenario under limited computational resources and processing time. The results of both cases are presented in Figure 10.

The improvement observed in Figure 10C and 10F, is directly attributable to the scalarization strategy defined in Equation (2). This formulation steers the optimization toward operating regions that simultaneously enhance both C-yield and C-SEC, rather than prioritizing one objective over the other. As a result, Bayesian optimization can effectively exploit synergistic directions in the design space where increased C-yield coincides with reduced C-SEC. This mechanism leads to a monotonic improvement of the composite objective during the early optimization stages, followed by stabilization once the Pareto-optimal region is reached. Importantly, the similar convergence patterns observed in both Case 1 and Case 2 indicate that the dominant improvement pathways are rapidly identified, even under strongly reduced data availability. This behavior highlights the effectiveness of the scalarized multi-objective formulation for accelerating convergence while preserving the essential structure of the Pareto front.

Figure 11 further illustrates the evolution of the optimization process in the objective space for both cases. The observed synergy between high C-yield and low C-SEC can be attributed to the intrinsic coupling between reaction conversion and recycle dynamics. As C-yield increases, a larger fraction of the reactants is converted within the reactor, thereby reducing the amount of unreacted gas circulating through the recycle loop. This reduction directly decreases compression and circulation requirements, which constitute a major contribution to the overall energy demand of the process. In addition, higher C-yield leads to increased MeAce production, which further reduces the specific energy cost when expressed per unit mass of product. These two effects act simultaneously and reinforce each other, resulting in a self-consistent improvement in both production and energy performance. This behavior reveals a key intensification mechanism in the DME carbonylation process, where improved conversion efficiency not only enhances production but also inherently reduces system-level energy consumption. Consequently, operating conditions that promote high C-yield inherently lead to improved C-SEC, demonstrating that the two objectives are positively coupled within the identified operating window rather than fundamentally conflicting.

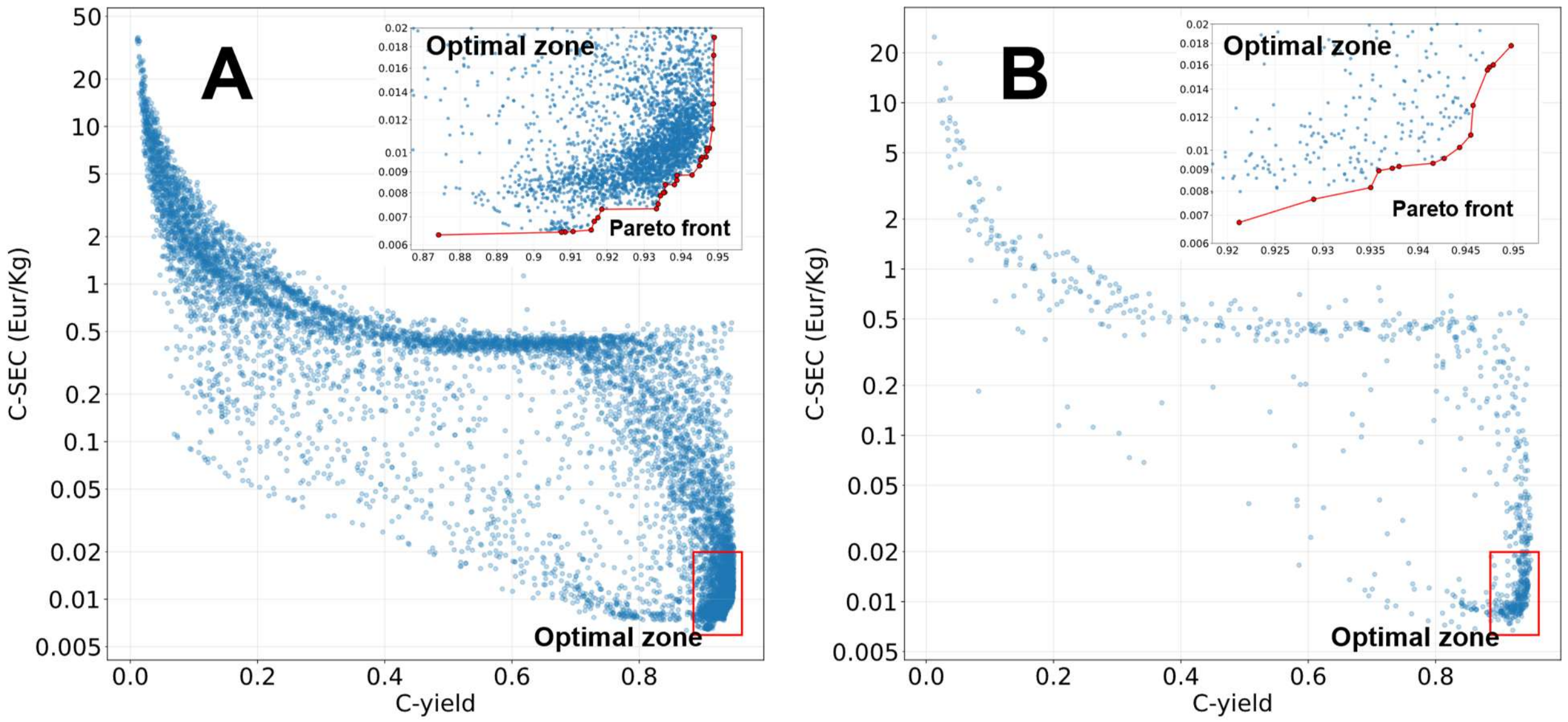

**Figure 11**. Distribution of BO-generated solutions in the methyl acetate –total energy objective space for Case 1 and Case 2.

Despite the substantial differences in initialization and optimization budget, both cases exhibit rapid convergence for both objectives. In particular, near-optimal values of C-yield and C-SEC are reached within approximately 200 BO iterations (see figure 10). Beyond this point, additional iterations or increased data availability do not lead to significant improvements in either objective. This observation indicates that the multi-objective BO framework is robust with respect to both data generation strategy and optimization budget for the investigated process, while reliably identifying the Pareto front that governs the trade-off between production performance and energy efficiency

The TPE is a widely used Bayesian optimization algorithm for efficient exploration of high-dimensional parameter spaces (Watanabe, 2025; Zhang et al., 2025). In this study, Bayesian optimization is implemented using the TPE algorithm, which employs a nonparametric density estimation approach to model the distributions of promising and less promising solutions directly in the parameter space. By partitioning observed objective values based on a quantile threshold and proposing new candidates through maximization of the ratio between the corresponding density functions, TPE achieves an effective balance between exploration and exploitation. Owing to its flexibility and robustness, TPE is particularly well suited for surrogate-assisted optimization of complex, nonlinear reaction–separation systems with strong variable interactions.

From a multi-objective perspective, the Pareto front represents the set of non-dominated solutions for which no further improvement in one objective can be achieved without compromising the other. In this study, the Pareto front captures the trade-off structure between C-yield and C-SEC and provides a comprehensive view of feasible optimal operating conditions. The convergence behavior observed in both Case 1 and Case 2 indicates that the optimization rapidly identifies and refines this Pareto-optimal region, even under strongly reduced data availability. Importantly, the similarity of the Pareto fronts obtained in both cases suggests that the essential trade-off structure of the process is effectively captured with limited initial data.

As the optimization progresses into the Pareto-optimal region, a clear trade-off between C-yield and C-SEC becomes evident. While earlier stages of the optimization reveal a favorable coupling between the two objectives, further improvement beyond this region requires increasingly significant energy penalties. In particular, increasing C-yield from approximately 91% to 94.5% is associated with an increase in C-SEC of about 50%. However, pushing the system toward even higher conversion levels reveals a much steeper trade-off: achieving a C-yield of around 95% requires nearly a doubling of C-SEC compared to the energy demand at 91% C-yield. This behavior highlights the nonlinear nature of the Pareto front, where incremental gains in production performance come at disproportionately higher energy costs as the system approaches its upper conversion limits. From a process intensification perspective, this finding provides a more comprehensive understanding of the trade-off structure between MeAce production efficiency and cost of energy consumption. Therefore, the Pareto front serves as an important decision-support tool, enabling the selection of appropriate operating conditions based on the balance between the benefits gained from increasing C-yield and the corresponding energy costs, depending on operational objectives and economic considerations of the process.

A comparison between the DOE and Bayesian optimization results highlights the effectiveness of the proposed framework. The DOE campaign (2730 simulations) achieved a maximum C-yield of ~88% with a C-SEC of ~0.012 EUR·kg$^{-1}$ MeAce. In contrast, Bayesian optimization improved C-yield to ~94.5% while reducing C-SEC to ~0.0065 EUR·kg$^{-1}$ MeAce. Despite the large DOE dataset, the BO framework increased C-yield by ~6.5% and nearly halved energy consumption. This demonstrates that surrogate-assisted optimization can identify superior operating regions beyond those accessible through extensive simulation campaigns, with significant implications for process design and industrial optimization.

To validate the reliability of the surrogate-assisted multi-objective Bayesian optimization results, process parameters of a subset of optimal zone were re-evaluated using direct steady-state process simulations established in the CoCo tool. From the set of Pareto-optimal operating points generated by BO, approximately 150 representative solutions covering the optimal zone were selected and simulated to assess the consistency between surrogate-based predictions and rigorous process calculations. The complete validation results for these points are provided in Appendix A, while a reduced set of 29 representative validation of pareto-font (see the red line in Figure 11A) is summarized in Table 4 for this validation.

The comparison between BO predictions and CoCo simulation results demonstrates strong agreement for both objectives. For C-yield, the relative deviation between surrogate-based predictions and process simulation results does not exceed 3% across all validated points. For C-SEC, the deviation remains within 10%, which is considered acceptable given the aggregated and system-level nature of the energy metric, which reflects coupled effects of compression, recycle flow, heat integration and pressure losses across multiple unit operations. Importantly, no systematic bias toward overestimation or underestimation is observed, indicating that the surrogate models preserve the simulation trends governing both production and energy consumption behavior in the Pareto-optimal region.

The re-simulated points confirm that high C-yield can be achieved within operating regimes characterized by relatively low C-SEC. This outcome substantiates the earlier observation from the DOE data analytics (see section 3.1) that production intensification and cost of energy consumption are not inherently antagonistic for the DME carbonylation process. Instead, favorable operating conditions simultaneously enhance conversion efficiency and reduce recycle-related compression demands, leading to a beneficial coupling between the two objectives.

**Table 4**. Validation of Pareto-optimal solutions from multi-objective Bayesian optimization using CoCo process simulations

| No | Reactor Settings | | | | | C-yield (%) | | C-SEC (Eur/Kg) | |
|---|---|---|---|---|---|---|---|---|---|
| | L | T | SR | P | CO | BO | CoCo | BO | CoCo |
| 1 | 8.80 | 486.0 | 0.18 | 3.03 | 45.22 | 94.9 | 95.0 | 0.0190 | 0.0094 |
| 2 | 8.80 | 486.0 | 0.18 | 2.99 | 45.31 | 94.9 | 94.9 | 0.0172 | 0.0093 |
| 3 | 8.80 | 486.0 | 0.17 | 3.03 | 45.42 | 94.9 | 94.8 | 0.0131 | 0.0092 |
| 4 | 8.80 | 486.0 | 0.15 | 2.98 | 45.52 | 94.9 | 94.7 | 0.0114 | 0.0091 |
| 5 | 8.80 | 486.0 | 0.14 | 2.96 | 45.82 | 94.8 | 94.5 | 0.0103 | 0.0090 |
| 6 | 8.73 | 486.0 | 0.14 | 3.02 | 45.56 | 94.7 | 94.7 | 0.0102 | 0.0090 |
| 7 | 8.80 | 486.0 | 0.15 | 2.97 | 45.98 | 94.7 | 94.3 | 0.0101 | 0.0090 |
| 8 | 8.80 | 486.0 | 0.14 | 3.03 | 45.81 | 94.7 | 94.5 | 0.0098 | 0.0089 |
| 9 | 8.80 | 486.0 | 0.15 | 2.97 | 46.21 | 94.6 | 94.1 | 0.0097 | 0.0089 |
| 10 | 8.80 | 486.0 | 0.15 | 2.84 | 46.55 | 94.5 | 93.7 | 0.0096 | 0.0089 |
| 11 | 8.73 | 485.9 | 0.13 | 2.92 | 46.16 | 94.5 | 94.1 | 0.0093 | 0.0089 |
| 12 | 8.80 | 485.1 | 0.13 | 3.00 | 46.26 | 94.3 | 94.0 | 0.0088 | 0.0089 |

| No | Reactor Settings | | | | | C-yield (%) | | C-SEC (Eur/Kg) | |
|---|---|---|---|---|---|---|---|---|---|
| | L | T | SR | P | CO | BO | CoCo | BO | CoCo |
| 13 | 8.80 | 484.2 | 0.13 | 2.93 | 46.70 | 93.9 | 93.5 | 0.0088 | 0.0090 |
| 14 | 8.68 | 484.7 | 0.12 | 2.99 | 46.28 | 93.9 | 93.9 | 0.0086 | 0.0089 |
| 15 | 8.80 | 484.9 | 0.13 | 2.92 | 47.21 | 93.8 | 93.1 | 0.0084 | 0.0088 |
| 16 | 8.63 | 484.2 | 0.11 | 3.18 | 46.25 | 93.6 | 94.1 | 0.0084 | 0.0089 |
| 17 | 8.80 | 485.0 | 0.12 | 3.00 | 47.30 | 93.6 | 93.1 | 0.0080 | 0.0088 |
| 18 | 8.66 | 484.5 | 0.12 | 2.91 | 47.24 | 93.5 | 93.0 | 0.0080 | 0.0088 |
| 19 | 8.59 | 484.9 | 0.11 | 3.07 | 46.90 | 93.4 | 93.5 | 0.0079 | 0.0088 |
| 20 | 8.65 | 484.5 | 0.09 | 3.18 | 46.49 | 93.4 | 93.9 | 0.0075 | 0.0086 |
| 21 | 8.64 | 484.5 | 0.09 | 3.06 | 46.65 | 93.3 | 93.6 | 0.0073 | 0.0085 |
| 22 | 8.50 | 482.1 | 0.08 | 3.15 | 46.78 | 91.8 | 93.3 | 0.0073 | 0.0088 |
| 23 | 8.39 | 482.3 | 0.08 | 3.15 | 47.81 | 91.7 | 92.3 | 0.0070 | 0.0087 |
| 24 | 8.50 | 482.0 | 0.08 | 3.20 | 47.98 | 91.6 | 92.2 | 0.0068 | 0.0086 |
| 25 | 8.42 | 483.4 | 0.07 | 3.21 | 48.28 | 91.6 | 92.1 | 0.0065 | 0.0082 |
| 26 | 8.42 | 483.6 | 0.07 | 3.22 | 48.96 | 91.1 | 91.5 | 0.0065 | 0.0081 |
| 27 | 8.41 | 483.5 | 0.07 | 3.21 | 49.18 | 90.9 | 91.2 | 0.0064 | 0.0080 |
| 28 | 8.42 | 483.5 | 0.07 | 3.21 | 49.35 | 90.8 | 91.1 | 0.0064 | 0.0080 |
| 29 | 8.80 | 486.0 | 0.11 | 3.60 | 53.07 | 87.4 | 89.0 | 0.0063 | 0.0069 |

*L: reactor length (m); T: reactor temperature (K); SR: splitter ratio; P: system pressure (MPa); CO: CO molar flow rate (mol/s)*

The strong agreement between surrogate-based optimization results and rigorous process simulations across a wide range of optimal conditions confirms the reliability of the multi-objective Bayesian optimization framework. Importantly, the surrogate models do not only perform well at the extreme points of the optimal region, but also accurately describe the intermediate trade-off zones along the Pareto front. The predicted Pareto structure reflects the true behavior of the process, rather than being an artifact of model extrapolation. From an engineering perspective, this consistency is critical for identifying reliable operating windows. It ensures that the optimal conditions suggested by the optimization remain valid when verified using detailed steady-state simulations. Therefore, the trade-offs identified between C-yield and C-SEC can be considered physically meaningful and can confidently support process intensification decisions as well as future experimental validation.

The multi-objective verification results reinforce the practical applicability of the proposed framework for energy-aware process intensification. By combining surrogate-assisted Bayesian optimization with targeted process simulation validation, the methodology enables efficient identification of operating windows that simultaneously maximize C-yield and minimize energy consumption. These findings further support the use of the proposed approach as a decision-support tool for early-stage process optimization and for guiding experimental or pilot-scale investigations toward energetically favorable and high-performance operating regimes.

### 3.4. Process insight and operating-region characterization for methyl acetate production

Additional insight into the multi-objective optimization results was obtained by examining the distribution of operating variables across different regions of the objective space using case 1. By projecting the BO-generated solutions onto the C-yield and C-SEC objective space and coloring them according to individual operating variables, distinct operating zones can be clearly identified, as illustrated in Figure 12.

The unfavorable operating zone, characterized by low C-yield and high C-SEC, is associated with relatively low reactor temperatures, short reactor lengths and low CO molar flow rates. In this region, insufficient reaction severity results in incomplete conversion, leading to high recycle flow rates and elevated compression energy requirements. Higher system pressures are also frequently observed, further contributing to increased C-SEC. No clear monotonic trend is observed for the splitter ratio in this zone, suggesting that its influence is either secondary or strongly coupled with other operating variables under low-conversion conditions.

In contrast, the favorable operating zone, defined by high C-yield combined with low C-SEC, exhibits a more compact and structured distribution of operating parameters. This region is characterized by elevated reactor temperatures (481- 486 K), longer reactor lengths (8.4 - 8.8 m), moderate CO molar flow rates (50 – 53 mol/s), and moderate system pressures (3.2 – 3.4 MPa). Under these conditions, higher conversion efficiency reduces recycling loads and compression requirements. In addition, the exothermic nature of the reaction leads to greater heat release, enhancing the potential for heat recovery and feed preheating through process heat integration, thereby improving overall energy performance. Compared to the unfavorable region, the favorable operating zone demonstrates greater robustness and controllability, making it particularly attractive for operation and scale-up.

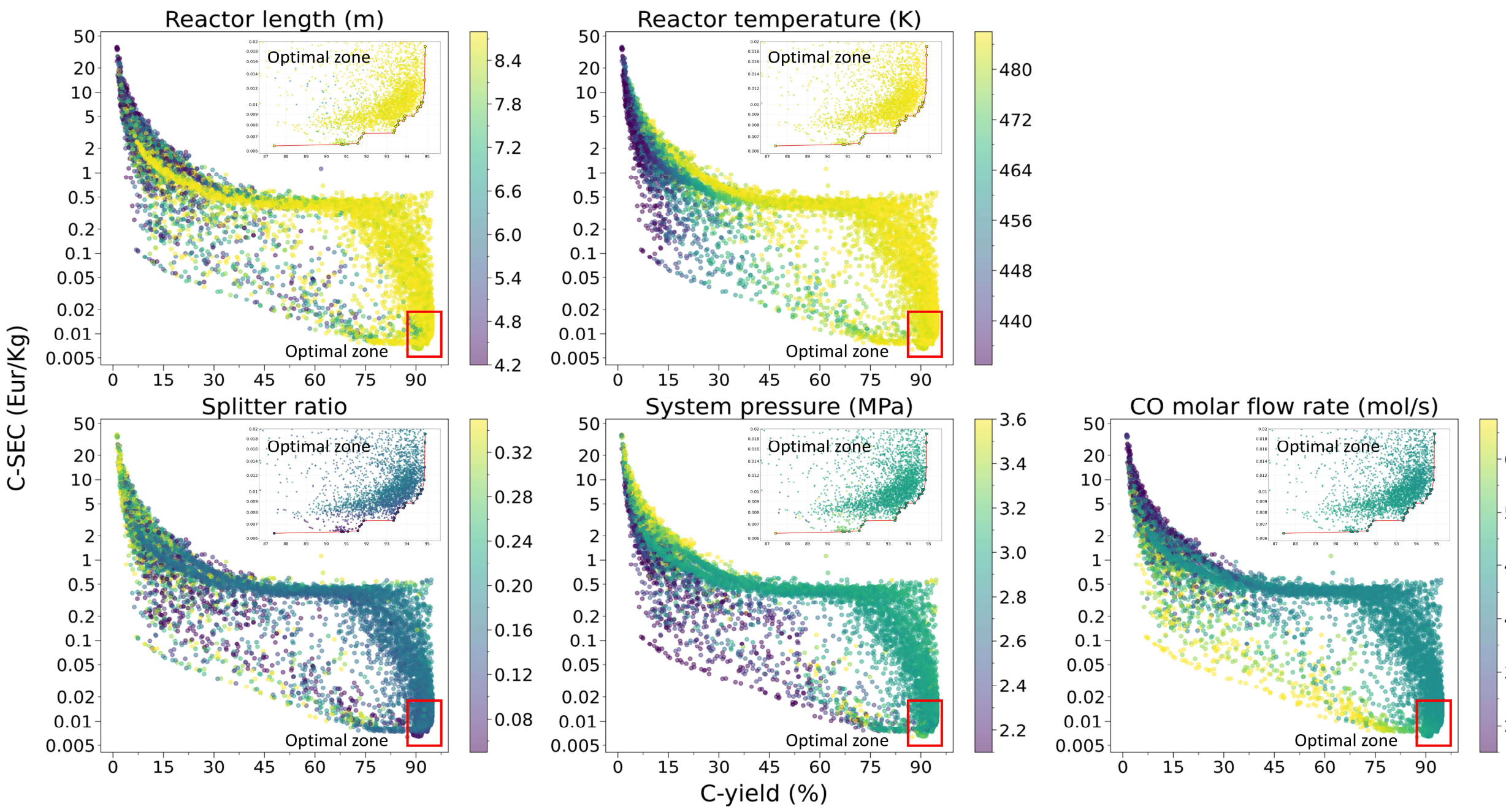

**Figure 12.** Distribution of operating parameters in the C-yield and C-SEC objective space obtained from multi-objective Bayesian optimization (Case 1)

The clear separation between favorable and unfavorable operating zones provides actionable insight into how operating variables jointly influence production and energy efficiency. These results further demonstrate the value of multi-objective Bayesian optimization not only for identifying optimal trade-offs but also for revealing physically meaningful operating regimes that can guide practical process design and operation. The analysis of operating variables in the unfavorable and favorable regions is consistent with the preliminary insights obtained from the DOE-based analysis of discretely sampled data presented in Section 3.1. The Pareto front region (Figure 12 and Table 4) indicates that individual parameters may vary only slightly; however, when combined with other factors, they can lead to significant changes in energy consumption cost. In particular, C-SEC may increase up to twofold at optimal points, while the corresponding improvement in yield remains relatively small.

From an engineering perspective, the existence of a well-defined favorable operating region provides actionable guidance for process design and operation. Rather than targeting a single optimal point, the proposed framework enables the selection of an operating window that balances performance, robustness, and operational flexibility. This is particularly relevant for industrial implementation, where uncertainties related to feed variability, catalyst aging, and operational disturbances necessitate conservative and stable operating strategies.

Furthermore, the rapid convergence of Bayesian optimization, combined with the consistency between surrogate-based predictions and rigorous process simulations, indicates that the proposed workflow can be effectively integrated into early-stage process development and debottlenecking studies. By reducing reliance on exhaustive simulation campaigns, the framework allows engineers to rapidly screen operating strategies, quantify trade-offs between competing objectives, and prioritize promising regions for experimental validation.

The results support the applicability of the surrogate-assisted multi-objective optimization framework as a practical decision-support tool for operating-window selection, process intensification, and energy-efficient operation of DME-to-methyl-acetate systems, while also providing a structured basis for subsequent pilot-scale testing and scale-up studies.

# 4 Conclusions

This study presents a data-driven framework for process intensification of C-yield and C-SEC via direct carbonylation of dimethyl ether and carbon monoxide. By integrating steady-state process simulation, surrogate modeling and Bayesian optimization the proposed approach enables systematic exploration and optimization of a highly nonlinear reaction–separation system with significantly reduced computational cost.

A comprehensive simulation-based dataset was generated to characterize the effects of key operating variables on C-yield and C-SEC. Machine learning surrogate models were developed and evaluated across a broad range of model families, demonstrating that advanced deep learning architectures provide superior accuracy and robustness, particularly under extrapolative conditions. The selected surrogate models were shown to reliably reproduce process simulation trends and were therefore suitable for optimization tasks.

Multi-objective optimization shows that high production rates can be achieved alongside reduced energy costs. Analysis of the optimal operating region indicates that higher conversion efficiency decreases recycling flow and compression demand while enhancing heat integration, leading to a favorable coupling between production intensification and energy efficiency. A clear and strongly nonlinear Pareto trade-off is observed between C-yield and C-SEC. In the moderate region, improvements in C-yield incur relatively small energy penalties, whereas near the upper conversion limit, further gains require disproportionately higher energy input. This highlights the practical limits of process intensification and emphasizes the importance of identifying optimal operating windows rather than maximizing conversion alone. The pareto front therefore serves as a key decision-support tool for balancing production performance and energy cost.

**Author Contributions**

Conceptualization, T.N.T., R.T.; Methodology, T.N.T., T.T.N., N.T., R.T.; Software, T.N.T., T.T.N.; Validation, T.N.T., T.T.N.; Formal Analysis, T.N.T., T.T.N.; Investigation, T.N.T., T.T.N.; Resources, T.N.T., N.T., R.T.; Data Curation: T.N.T., T.T.N.; Writing – Original Draft Preparation, T.N.T., T.T.N., N.T., R.T.; Writing – Review & Editing, T.N.T., T.T.N., N.T., R.T.; Visualization, T.N.T., T.T.N.; Supervision, T.N.T; Project Administration, T.N.T., N.T., R.T.; Funding

Acquisition, T.N.T., N.T., R.T. All authors have read and agreed to the published version of the manuscript.

**Funding**

This research was supported by the Strategic Institute Programme - SINTEF Industry

**Data Availability Statement**

The original data presented in the study are openly available in Surrogate-based multi-objective optimization for methyl acetate process intensification at https://gitlab.sintef.no/trung.trinh

**Acknowledgments**

The authors also thank Torbjørn Pettersen for valuable discussions and technical support related to pinch analysis and heat integration.

**Appendix A.**

Supplementary data.

**Declaration of competing interests**

The authors declare that they have no known competing financial interests or personal relationships that could have appeared to influence the work reported in this paper.

Appendix A.

Appendix 1: MLP model structure

1. DualHeadResNet

The proposed DualHeadResNet is a residual multilayer perceptron for regression tasks. The input features are first projected into a latent representation, followed by a stack of residual feed-forward blocks. Each block consists of layer normalization, a nonlinear transformation, and a skip connection, where the block output is added to its input. This residual structure supports gradient propagation in deeper architectures.

The model produces its output through two parallel components: a base head and a tail head. Both heads operate on the shared latent representation. The base head generates a scalar output that is passed through a sigmoid activation and scaled by a learnable parameter, resulting in a bounded baseline prediction. The tail head produces an additional scalar output, which is transformed using a softplus function to ensure non-negativity.

The contribution of the tail component is controlled by a learnable scalar parameter passed through a sigmoid function, acting as a global gating factor. The final prediction is computed as the sum of the scaled base output and the gated tail component. This formulation allows the model to represent the target as a combination of a bounded baseline term and a non-negative additive adjustment.

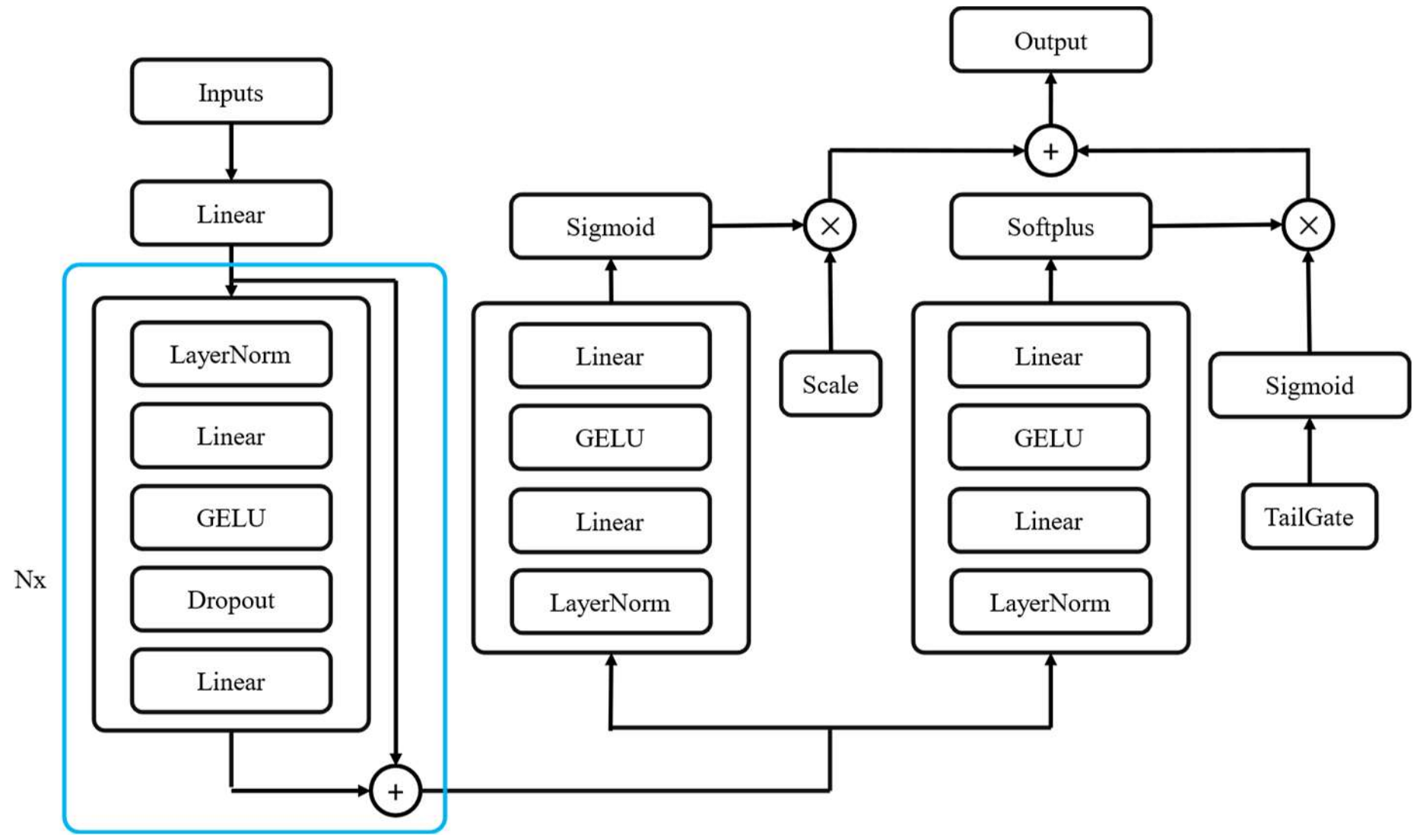

Fig. A.1. Architecture of the DualHeadResNet model

2. Residual MLP

The ResidualMLP is a deep residual multilayer perceptron designed as a strong baseline for regression tasks. The input features are first linearly projected into a fixed-dimensional latent space, enabling subsequent nonlinear transformations to operate in a shared representation domain.

The core of the model consists of multiple residual feed-forward blocks, each composed of layer normalization, an expansion–compression linear transformation, nonlinear activation, and dropout regularization. Residual (skip) connections are employed to preserve information flow across layers and to mitigate gradient degradation, thereby improving optimization stability and convergence behavior in deeper networks.

A single output head is applied to the final latent representation to produce the regression output. This head performs additional nonlinear feature transformation followed by a linear mapping to a scalar prediction. Unlike DualHeadResNet, ResidualMLP employs a unified output pathway without explicit decomposition, allowing the model to learn a global mapping from inputs to outputs in an end-to-end manner.

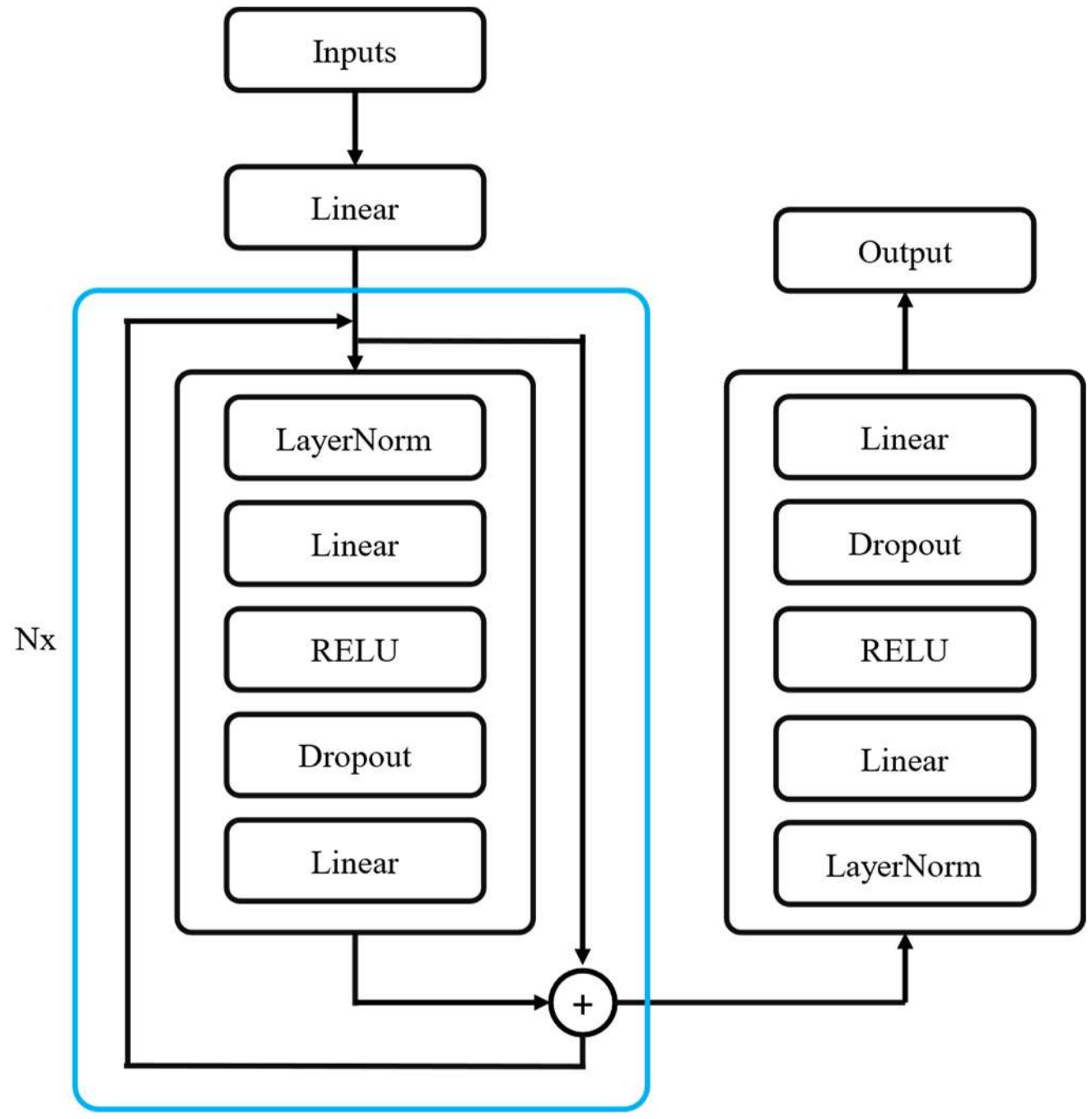

Fig. A.2. Architecture of the ResidualMLP model

3. MLPEnsemble

The MLPEnsemble model adopts an ensemble-based regression strategy, in which multiple deep multilayer perceptrons are trained jointly and their predictions are aggregated to improve robustness and generalization. The input features are first processed by a shared feature transformation module that maps the original input space into a compact latent representation through successive linear transformations, nonlinear activations, and dropout regularization.

Following the shared feature extractor, the transformed representation is fed in parallel into multiple independent MLP branches, referred to as trees. Each branch consists of a stack of fully connected layers with repeated nonlinear transformations and batch normalization, enabling the ensemble members to learn diverse nonlinear mappings over the same feature space. The final prediction of the model is obtained by averaging the outputs of all branches, which effectively reduces variance and mitigates sensitivity to local overfitting.

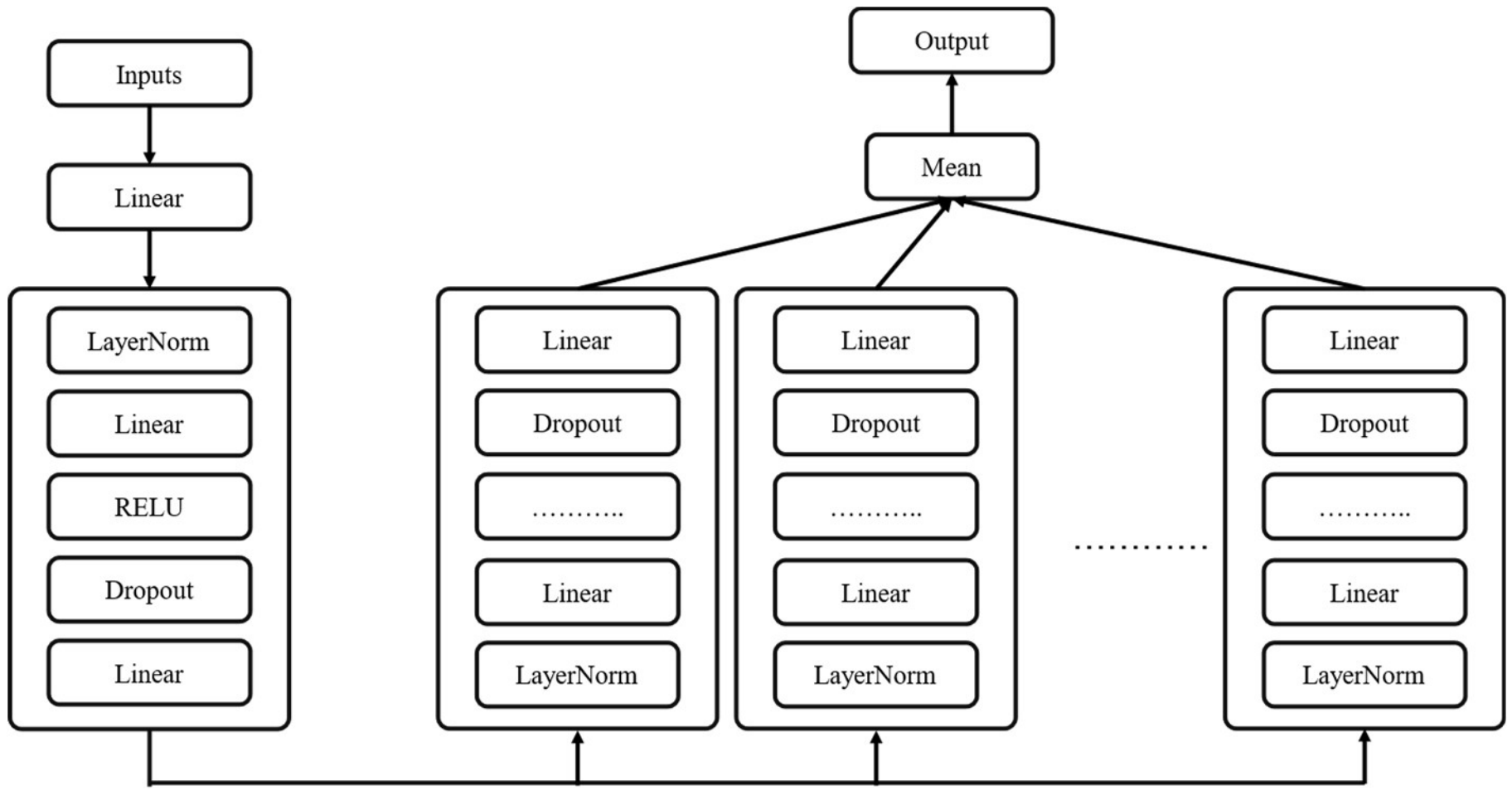

Fig. A.3. Architecture of the MLPEnsemble model

4. MediumDNN

MediumDNN is a feed-forward neural network with moderate depth and capacity, suitable for regression tasks involving more complex feature interactions. The architecture comprises three fully connected hidden layers with 128, 64, and 32 neurons, respectively, each followed by ReLU activation to model non-linear patterns. Dropout layers with a rate of 0.2 are applied after the first two hidden layers to mitigate overfitting and enhance generalization. The model concludes with a single linear output neuron, enabling continuous value prediction while offering a balance between expressive power and computational efficiency.

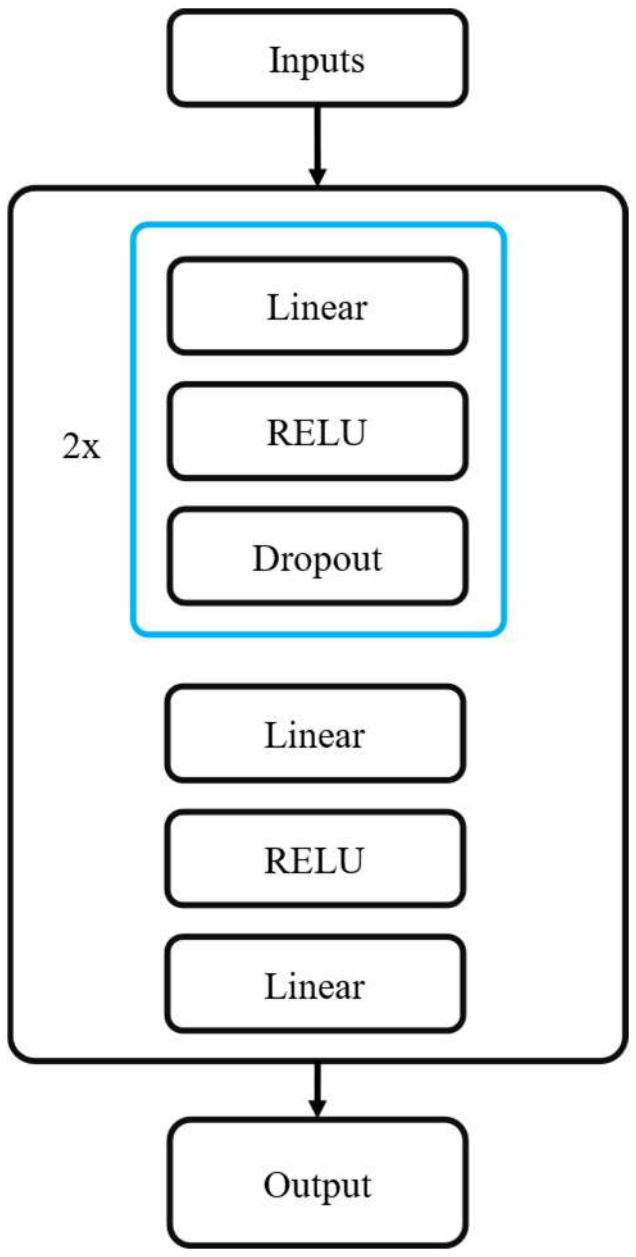

Fig. A.4. Architecture of the MediumDNN model

5. DeepNet

DeepNet is a deep feed-forward neural network designed to model complex, high-dimensional relationships in regression tasks. The architecture consists of four fully connected hidden layers with decreasing widths of 256, 128, 64, and 32 neurons, each followed by ReLU activation to introduce non-linearity. Dropout regularization is applied after the first three hidden layers with rates of 0.3, 0.2, and 0.2, respectively, to reduce overfitting as model depth increases. The network ends with a single linear output neuron, enabling accurate continuous value prediction while providing high representational capacity.

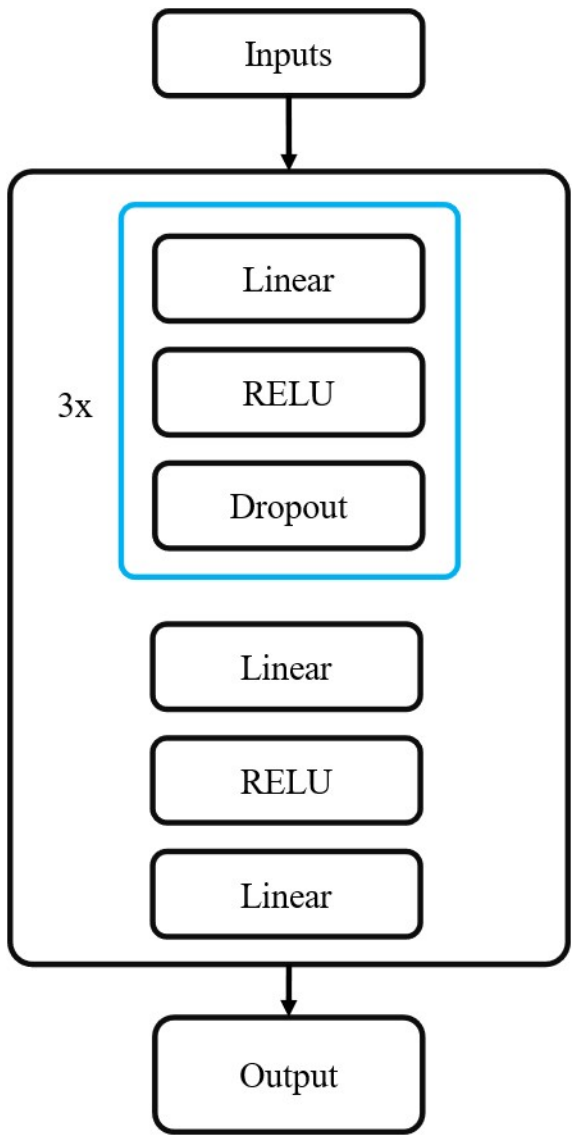

Fig. A.5. Architecture of the DeepNet model

Appendix 2. Model validation

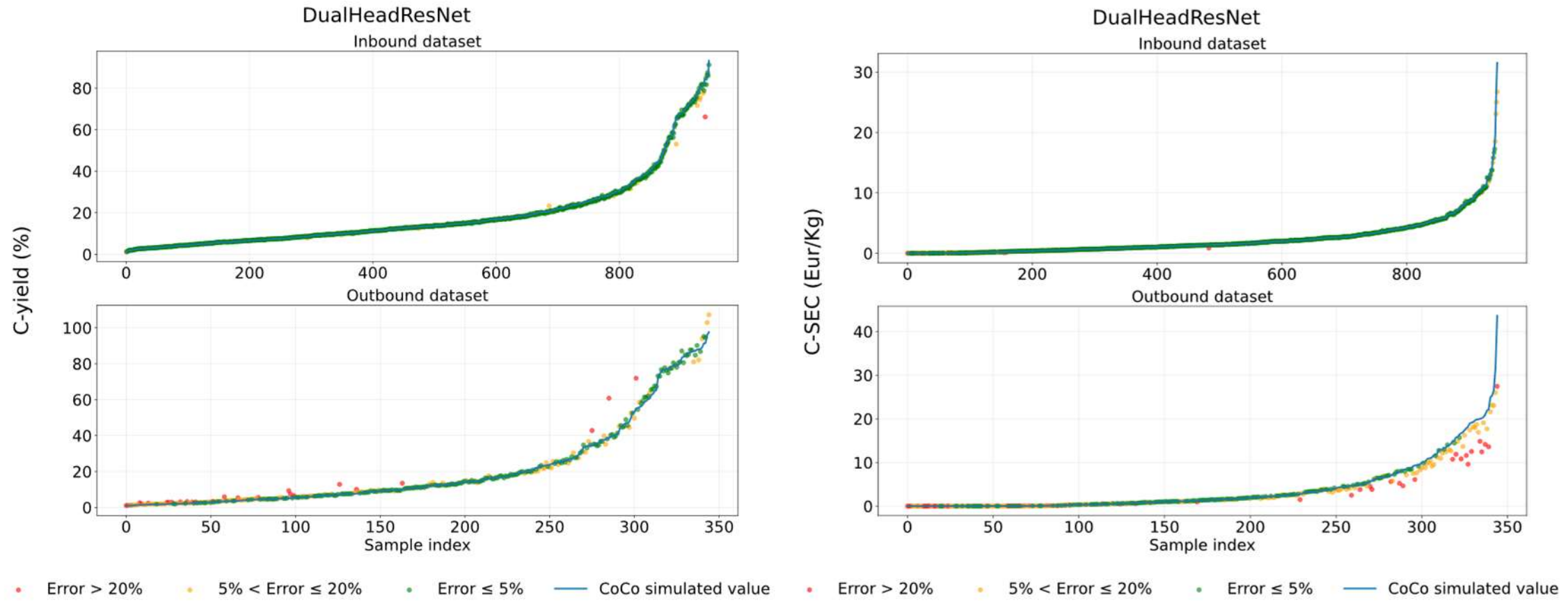

Fig. A.6. Validation of DualHeadResNet model on in-domain and out-of-domain datasets for c-yield and C-SEC

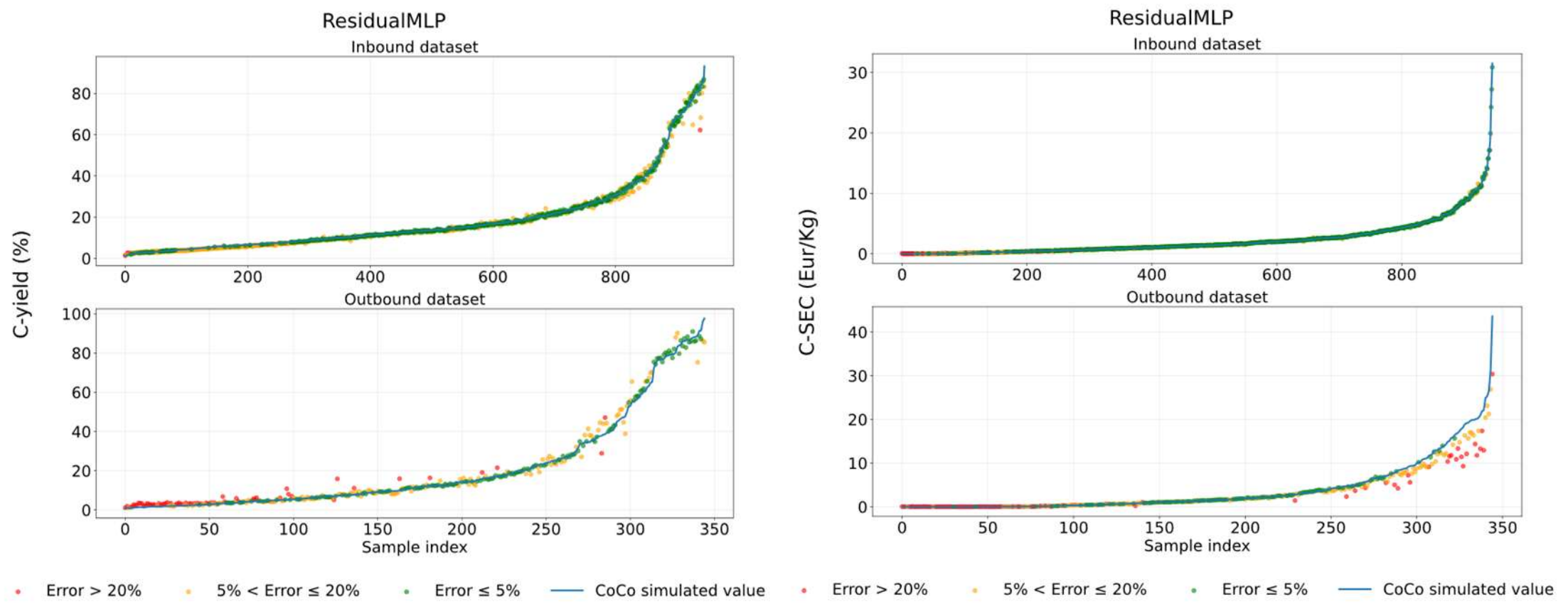

Fig. A.7. Validation of ResidualMLP model on in-domain and out-of-domain datasets for c-yield and C-SEC

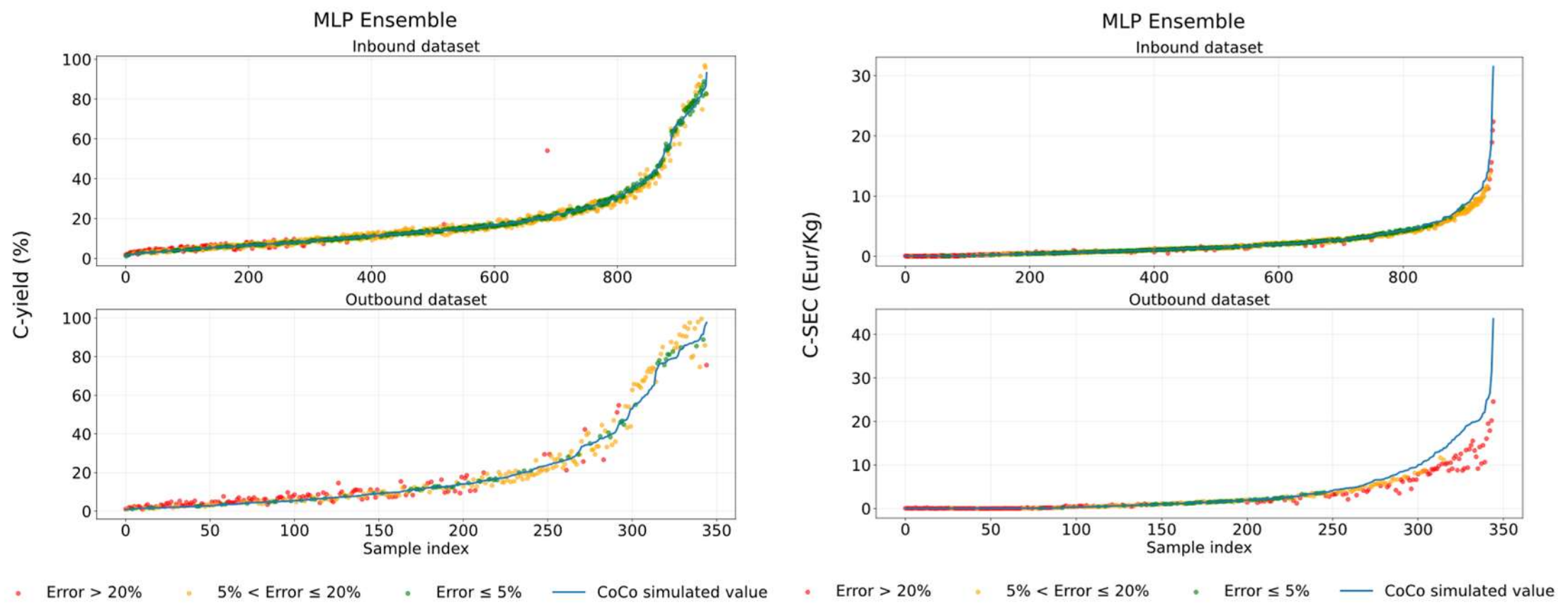

Fig. A.8. Validation of MLP Ensemble model on in-domain and out-of-domain datasets for c-yield and C-SEC

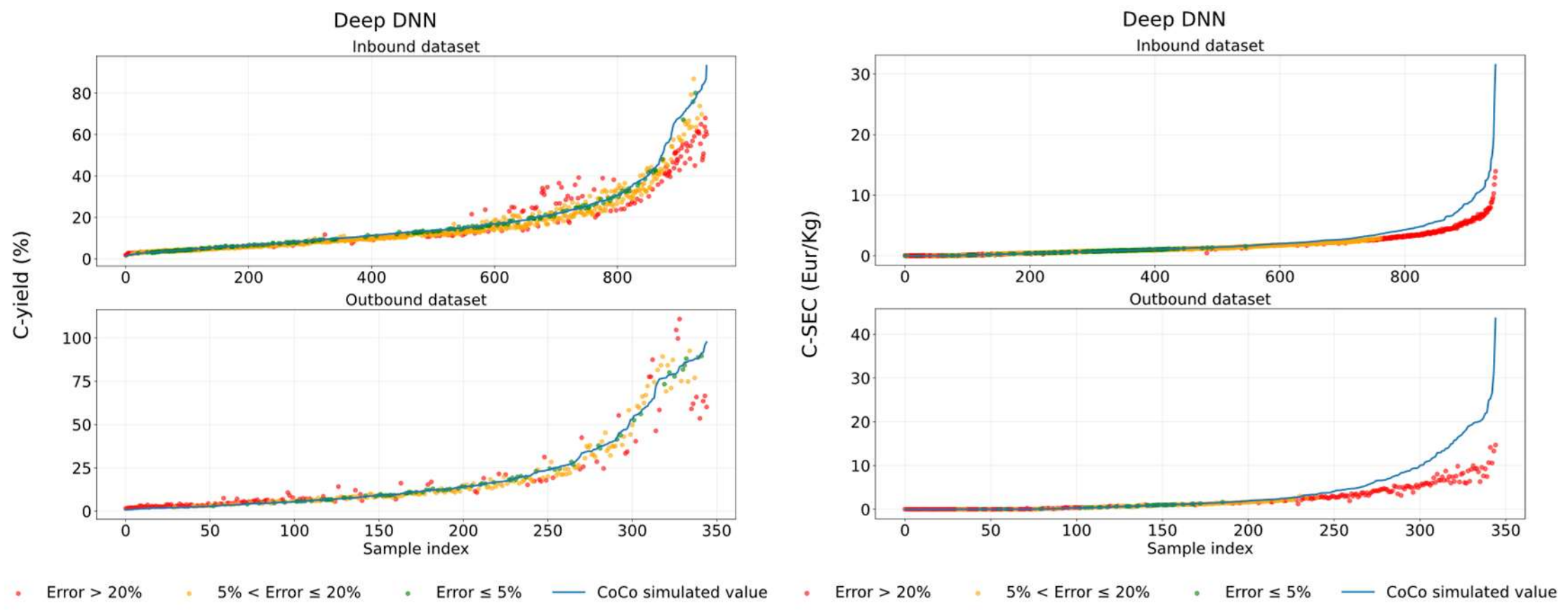

Fig. A.9. Validation of Deep DNN model on in-domain and out-of-domain datasets for c-yield and C-SEC

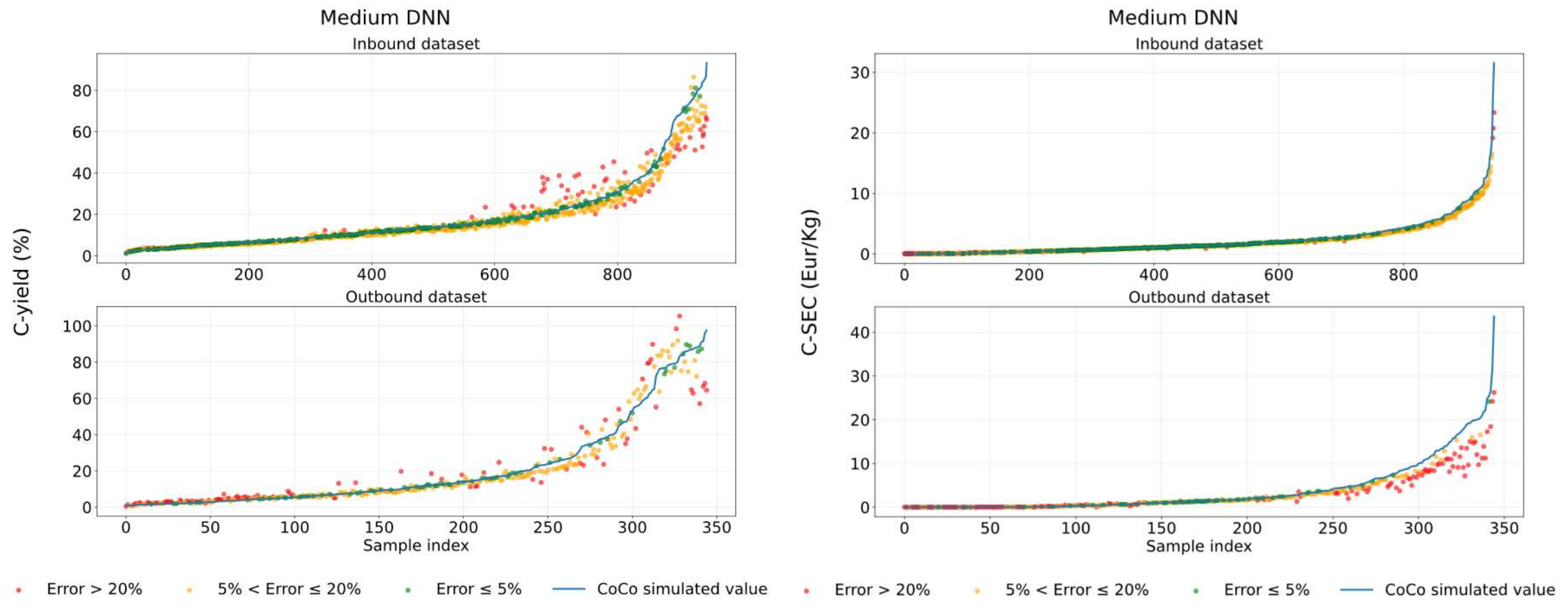

Fig. A.10. Validation of Medium DNN model on in-domain and out-of-domain datasets for c-yield and C-SEC

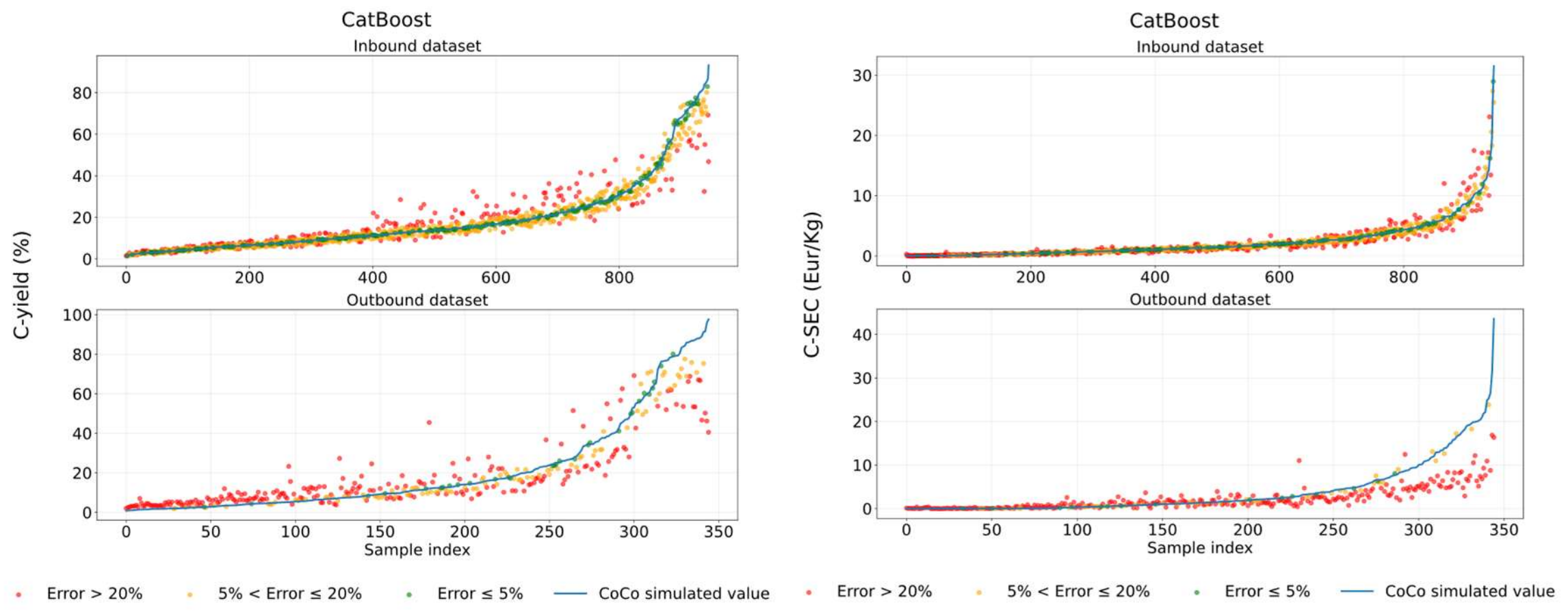

Fig. A.11. Validation of CatBoot model on in-domain and out-of-domain datasets for c-yield and C-SEC

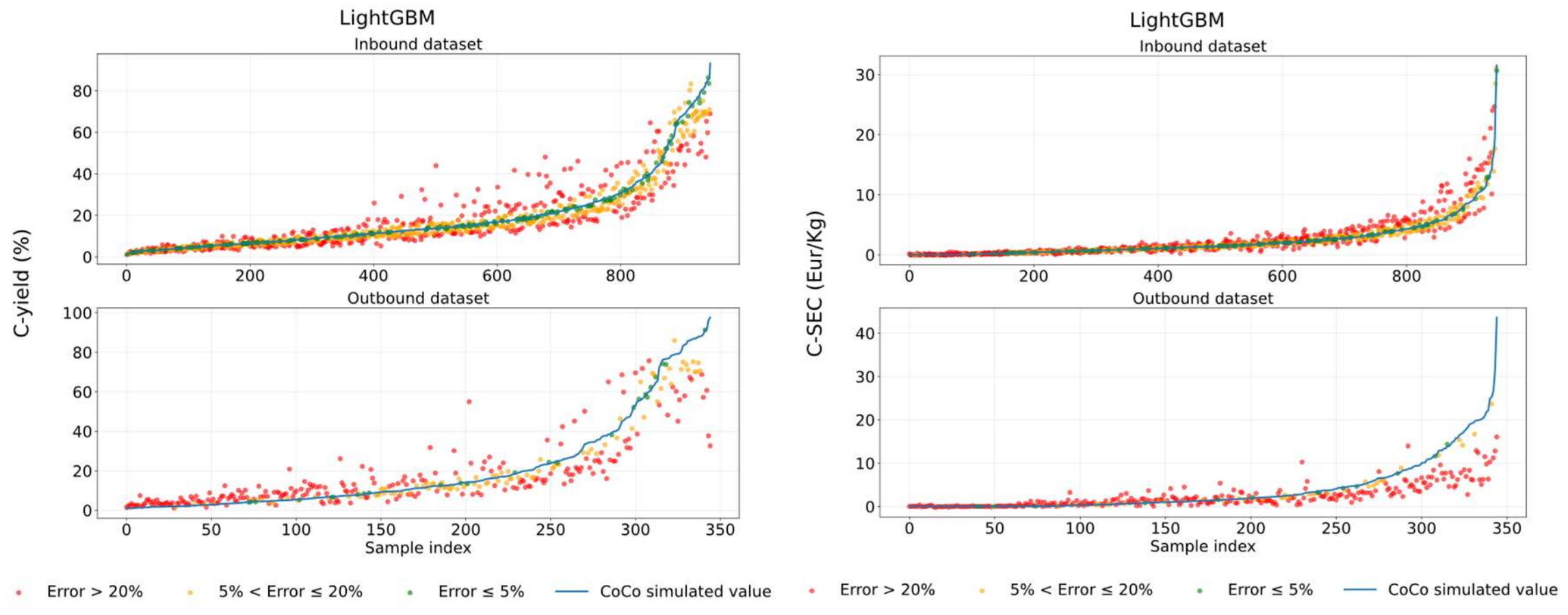

Fig. A.12. Validation of LightGBM model on in-domain and out-of-domain datasets for c-yield and C-SEC

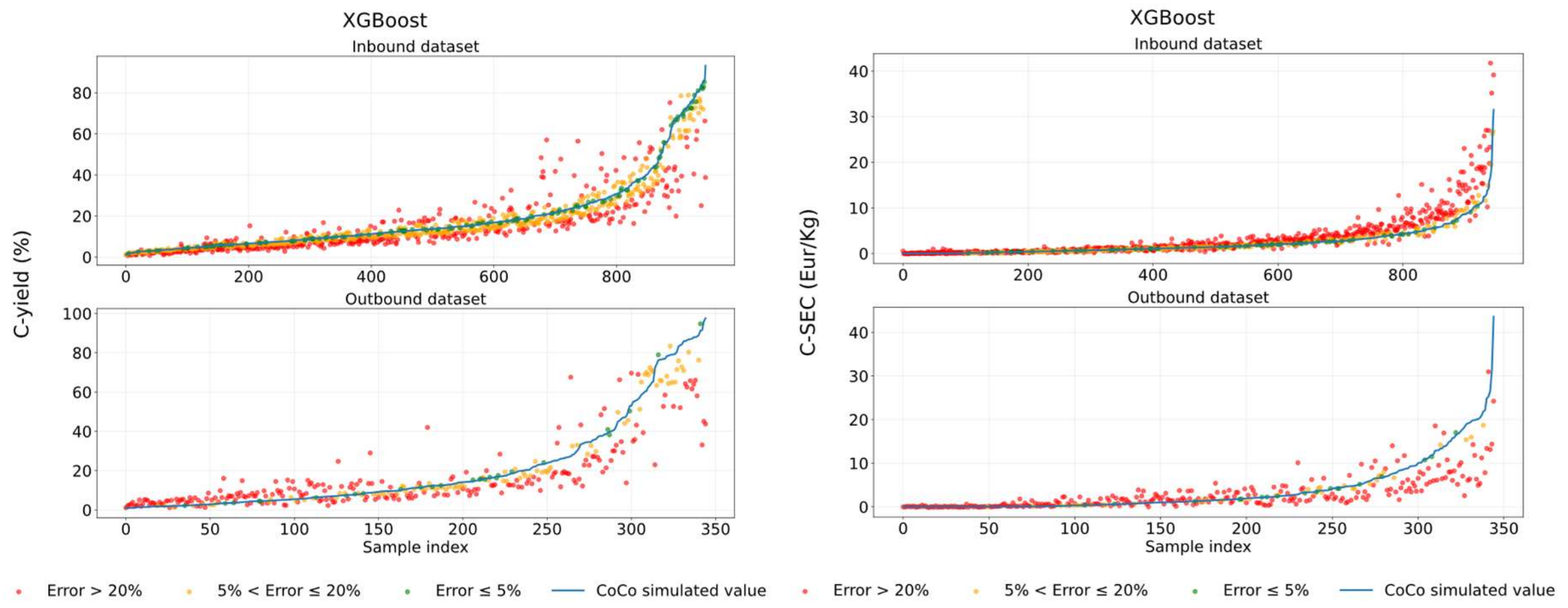

Fig. A.13. Validation of XGBoost model on in-domain and out-of-domain datasets for c-yield and C-SEC

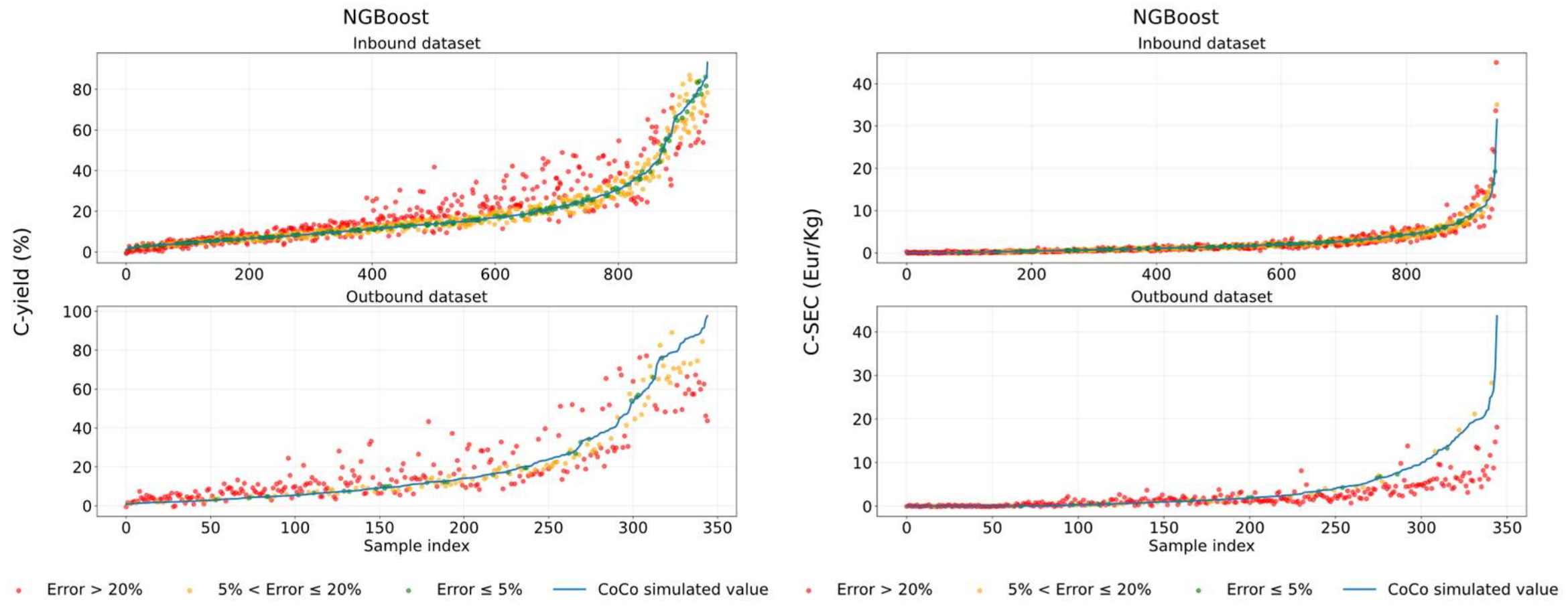

Fig. A.14. Validation of NGBoost model on in-domain and out-of-domain datasets for c-yield and C-SEC

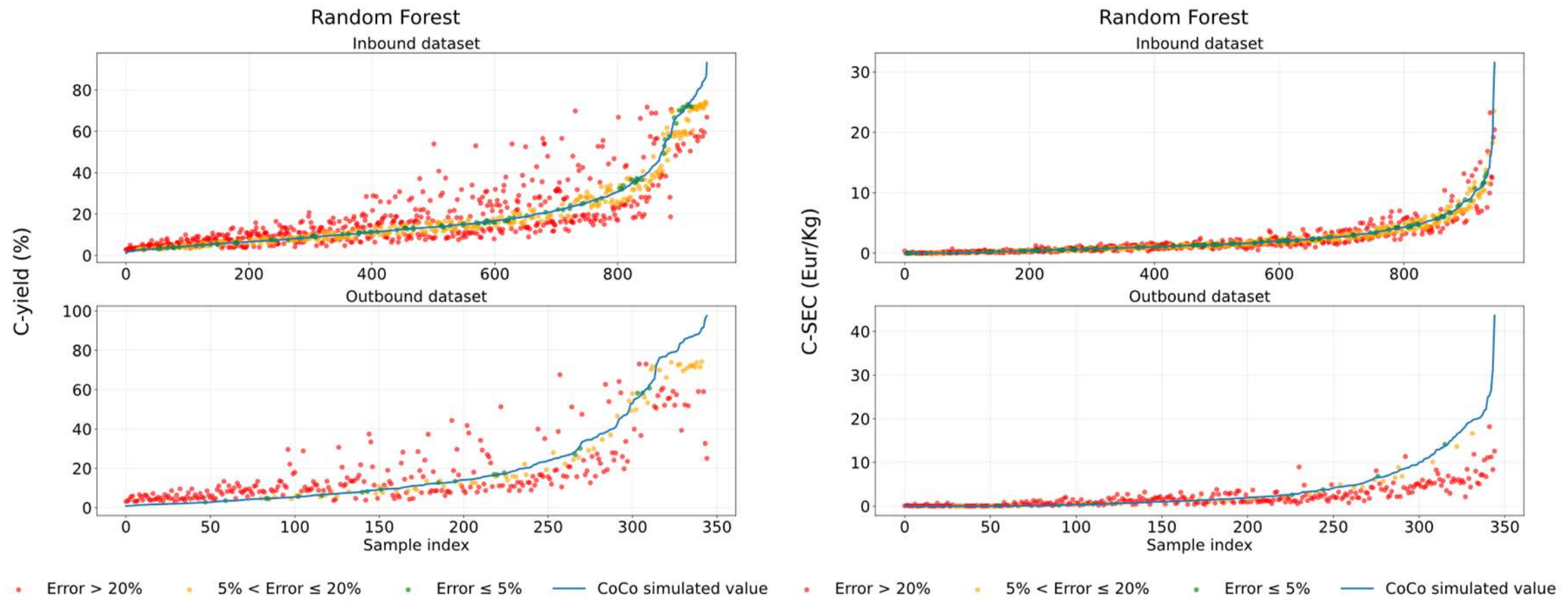

Fig. A.15. Validation of Random Forest model on in-domain and out-of-domain datasets for c-yield and C-SEC

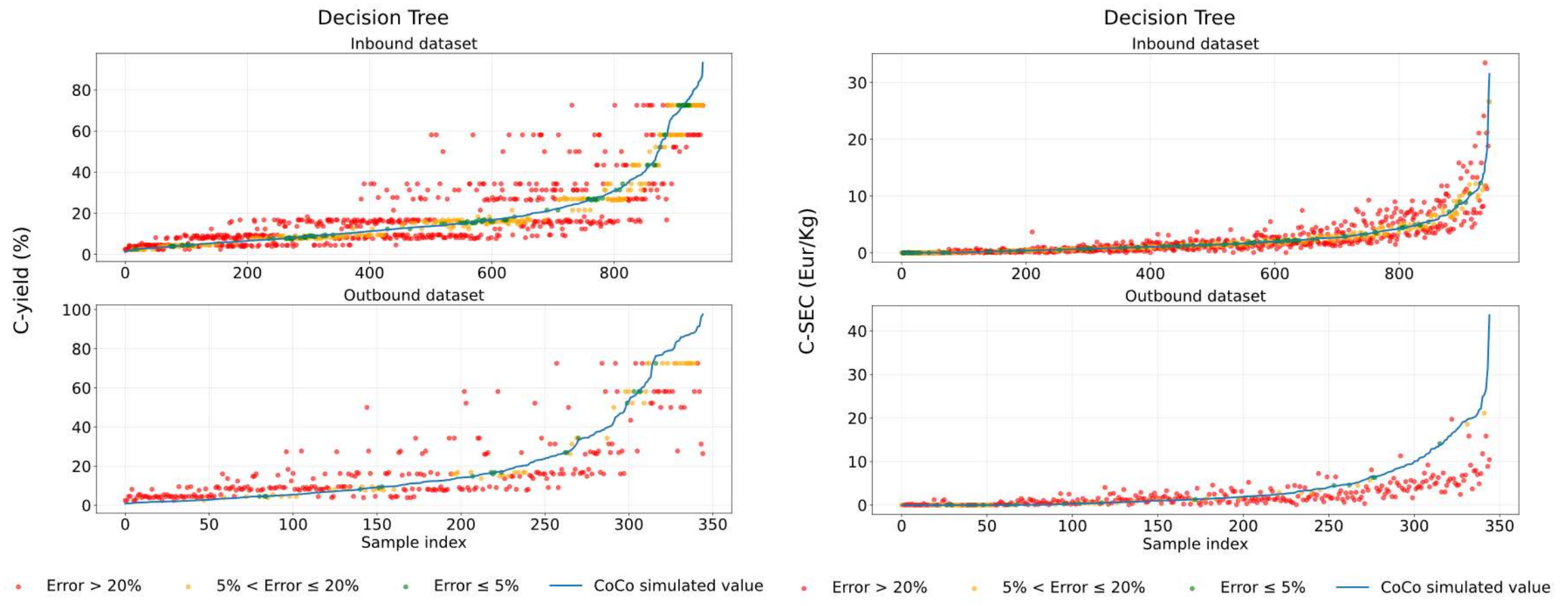

Fig. A.16. Validation of Decision Tree model on in-domain and out-of-domain datasets for c-yield and C-SEC

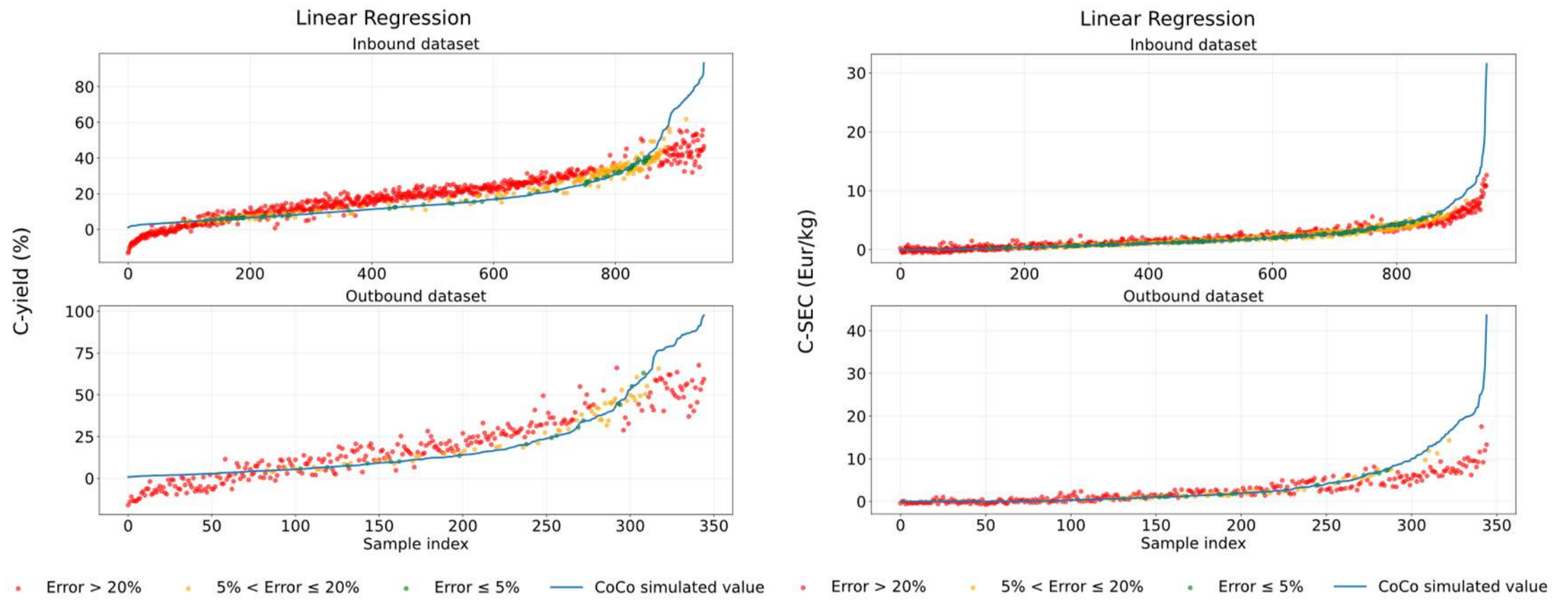

Fig. A.17. Validation of Linear Regression model on in-domain and out-of-domain datasets for c-yield and C-SEC

**Appendix 3.** Bayesian optimalization results of multiple targets

Table A.1. Validation of Bayesian optimization results for multiple targets

| No | Reactor Settings | | | | | C-yield (%) | | C-SEC (Eur/kg) | |
|---|---|---|---|---|---|---|---|---|---|
| | **L** | **T** | **SR** | **P** | **CO** | **BO** | **CoCo** | **BO** | **CoCo** |
| 1 | 8.72 | 486.0 | 0.14 | 2.96 | 46.07 | 94.6 | 94.2 | 0.0098 | 0.0090 |
| 2 | 8.80 | 486.0 | 0.15 | 2.88 | 46.63 | 94.5 | 93.7 | 0.0098 | 0.0090 |
| 3 | 8.80 | 485.0 | 0.16 | 2.89 | 46.57 | 94.3 | 93.6 | 0.0100 | 0.0091 |
| 4 | 8.72 | 486.0 | 0.15 | 2.91 | 46.79 | 94.2 | 93.5 | 0.0093 | 0.0089 |
| 5 | 8.68 | 484.8 | 0.13 | 3.05 | 46.03 | 94.2 | 94.2 | 0.0094 | 0.0091 |
| 6 | 8.71 | 486.0 | 0.15 | 2.97 | 46.60 | 94.1 | 93.8 | 0.0095 | 0.0090 |
| 7 | 8.72 | 486.0 | 0.15 | 3.02 | 46.43 | 94.0 | 94.0 | 0.0097 | 0.0090 |
| 8 | 8.72 | 485.1 | 0.15 | 3.01 | 46.50 | 94.0 | 93.8 | 0.0096 | 0.0091 |
| 9 | 8.57 | 485.0 | 0.11 | 3.06 | 45.82 | 94.0 | 94.4 | 0.0090 | 0.0089 |
| 10 | 8.80 | 484.4 | 0.15 | 3.05 | 46.43 | 94.0 | 93.9 | 0.0099 | 0.0092 |
| 11 | 8.63 | 484.3 | 0.15 | 2.96 | 46.45 | 93.9 | 93.7 | 0.0098 | 0.0093 |
| 12 | 8.67 | 484.7 | 0.13 | 2.94 | 46.13 | 93.9 | 94.0 | 0.0096 | 0.0091 |
| 13 | 8.65 | 484.9 | 0.16 | 2.88 | 47.00 | 93.9 | 93.2 | 0.0098 | 0.0092 |
| 14 | 8.71 | 484.2 | 0.15 | 2.95 | 46.73 | 93.9 | 93.4 | 0.0095 | 0.0092 |
| 15 | 8.66 | 485.0 | 0.13 | 2.97 | 46.83 | 93.8 | 93.5 | 0.0087 | 0.0090 |
| 16 | 8.66 | 484.3 | 0.12 | 2.99 | 46.39 | 93.8 | 93.8 | 0.0088 | 0.0090 |
| 17 | 8.53 | 485.0 | 0.13 | 2.90 | 46.68 | 93.8 | 93.4 | 0.0089 | 0.0090 |
| 18 | 8.67 | 484.5 | 0.13 | 2.96 | 46.18 | 93.8 | 94.0 | 0.0093 | 0.0091 |
| 19 | 8.65 | 485.0 | 0.16 | 2.95 | 46.83 | 93.8 | 93.4 | 0.0098 | 0.0092 |
| 20 | 8.35 | 486.0 | 0.13 | 2.91 | 46.75 | 93.8 | 93.4 | 0.0087 | 0.0089 |
| 21 | 8.52 | 485.1 | 0.13 | 2.98 | 46.65 | 93.8 | 93.6 | 0.0087 | 0.0090 |
| 22 | 8.73 | 485.1 | 0.15 | 3.03 | 46.69 | 93.8 | 93.7 | 0.0093 | 0.0091 |
| 23 | 8.52 | 485.2 | 0.14 | 2.99 | 46.72 | 93.7 | 93.5 | 0.0092 | 0.0091 |
| 24 | 8.68 | 484.7 | 0.12 | 2.96 | 46.11 | 93.7 | 94.0 | 0.0088 | 0.0089 |
| 25 | 8.62 | 484.2 | 0.14 | 2.99 | 46.74 | 93.7 | 93.5 | 0.0091 | 0.0092 |
| 26 | 8.80 | 483.2 | 0.15 | 3.01 | 46.82 | 93.7 | 93.4 | 0.0095 | 0.0093 |
| 27 | 8.80 | 486.0 | 0.17 | 3.00 | 46.94 | 93.6 | 93.5 | 0.0097 | 0.0090 |
| 28 | 8.72 | 484.3 | 0.16 | 2.94 | 47.10 | 93.6 | 93.1 | 0.0097 | 0.0093 |
| 29 | 8.66 | 485.9 | 0.17 | 2.81 | 47.68 | 93.6 | 92.6 | 0.0097 | 0.0090 |
| 30 | 8.72 | 485.1 | 0.14 | 2.94 | 47.29 | 93.6 | 93.1 | 0.0087 | 0.0089 |
| 31 | 8.60 | 483.7 | 0.12 | 3.06 | 46.00 | 93.6 | 94.1 | 0.0094 | 0.0092 |
| 32 | 8.71 | 486.0 | 0.17 | 2.96 | 47.06 | 93.6 | 93.4 | 0.0098 | 0.0090 |
| 33 | 8.63 | 484.2 | 0.11 | 3.19 | 46.25 | 93.6 | 94.1 | 0.0084 | 0.0089 |
| 34 | 8.68 | 484.6 | 0.12 | 2.94 | 47.25 | 93.5 | 93.0 | 0.0080 | 0.0088 |
| 35 | 8.65 | 486.0 | 0.17 | 2.89 | 47.34 | 93.5 | 93.0 | 0.0099 | 0.0091 |
| 36 | 8.72 | 483.6 | 0.13 | 2.93 | 46.93 | 93.5 | 93.2 | 0.0088 | 0.0091 |
| 37 | 8.64 | 484.1 | 0.16 | 2.99 | 46.88 | 93.5 | 93.3 | 0.0098 | 0.0093 |
| 38 | 8.71 | 484.1 | 0.15 | 2.89 | 47.40 | 93.5 | 92.8 | 0.0092 | 0.0092 |
| 39 | 8.65 | 484.3 | 0.12 | 3.18 | 46.36 | 93.5 | 94.0 | 0.0087 | 0.0090 |
| 40 | 8.65 | 485.0 | 0.17 | 2.91 | 47.31 | 93.5 | 93.0 | 0.0096 | 0.0092 |
| 41 | 8.72 | 486.0 | 0.17 | 2.87 | 47.62 | 93.4 | 92.8 | 0.0094 | 0.0090 |
| 42 | 8.72 | 483.5 | 0.16 | 2.96 | 47.14 | 93.4 | 93.0 | 0.0097 | 0.0094 |
| 43 | 8.60 | 486.0 | 0.14 | 2.93 | 47.45 | 93.4 | 93.0 | 0.0084 | 0.0088 |
| 44 | 8.65 | 486.0 | 0.15 | 3.04 | 46.98 | 93.4 | 93.5 | 0.0089 | 0.0089 |
| 45 | 8.65 | 483.6 | 0.14 | 2.93 | 46.45 | 93.4 | 93.6 | 0.0099 | 0.0094 |
| 46 | 8.64 | 485.0 | 0.16 | 3.04 | 46.75 | 93.4 | 93.6 | 0.0099 | 0.0092 |
| 47 | 8.66 | 484.3 | 0.09 | 3.18 | 46.43 | 93.4 | 93.9 | 0.0075 | 0.0086 |
| 48 | 8.59 | 483.7 | 0.12 | 3.00 | 46.99 | 93.4 | 93.2 | 0.0083 | 0.0091 |
| 49 | 8.51 | 484.9 | 0.14 | 3.01 | 46.95 | 93.4 | 93.3 | 0.0089 | 0.0091 |
| 50 | 8.68 | 484.7 | 0.12 | 3.06 | 47.16 | 93.3 | 93.2 | 0.0079 | 0.0088 |
| 51 | 8.71 | 483.3 | 0.16 | 2.86 | 47.46 | 93.3 | 92.5 | 0.0098 | 0.0094 |
| 52 | 8.64 | 484.1 | 0.17 | 2.77 | 47.58 | 93.3 | 92.3 | 0.0099 | 0.0094 |
| 53 | 8.65 | 484.2 | 0.13 | 2.94 | 47.48 | 93.3 | 92.8 | 0.0082 | 0.0089 |
| 54 | 8.73 | 484.3 | 0.14 | 3.01 | 47.27 | 93.3 | 93.1 | 0.0086 | 0.0091 |
| 55 | 8.72 | 484.3 | 0.16 | 2.87 | 47.75 | 93.3 | 92.5 | 0.0091 | 0.0091 |
| 56 | 8.69 | 483.8 | 0.12 | 3.02 | 45.95 | 93.2 | 94.2 | 0.0092 | 0.0091 |
| 57 | 8.72 | 485.1 | 0.16 | 3.00 | 47.21 | 93.2 | 93.2 | 0.0093 | 0.0091 |
| 58 | 8.65 | 484.1 | 0.09 | 3.05 | 46.50 | 93.2 | 93.7 | 0.0075 | 0.0086 |
| 59 | 8.57 | 484.9 | 0.12 | 3.01 | 47.31 | 93.2 | 93.1 | 0.0081 | 0.0088 |
| 60 | 8.56 | 486.0 | 0.14 | 3.03 | 47.17 | 93.2 | 93.3 | 0.0086 | 0.0089 |
| 61 | 8.65 | 486.0 | 0.16 | 3.04 | 46.91 | 93.2 | 93.6 | 0.0096 | 0.0091 |
| 62 | 8.45 | 484.4 | 0.15 | 2.97 | 47.09 | 93.2 | 93.1 | 0.0094 | 0.0093 |
| 63 | 8.80 | 485.0 | 0.15 | 3.08 | 47.03 | 93.2 | 93.5 | 0.0091 | 0.0090 |
| 64 | 8.65 | 485.1 | 0.15 | 3.03 | 47.17 | 93.2 | 93.3 | 0.0088 | 0.0090 |
| 65 | 8.49 | 483.4 | 0.15 | 3.01 | 46.88 | 93.2 | 93.2 | 0.0098 | 0.0095 |
| 66 | 8.73 | 485.1 | 0.17 | 2.80 | 48.15 | 93.2 | 92.1 | 0.0093 | 0.0091 |
| 67 | 8.75 | 486.0 | 0.17 | 2.80 | 48.20 | 93.2 | 92.2 | 0.0092 | 0.0089 |
| 68 | 8.65 | 482.7 | 0.15 | 3.03 | 47.06 | 93.1 | 93.1 | 0.0092 | 0.0094 |
| 69 | 8.51 | 485.1 | 0.16 | 2.95 | 47.29 | 93.1 | 93.0 | 0.0095 | 0.0092 |
| 70 | 8.66 | 484.9 | 0.13 | 2.69 | 48.14 | 93.1 | 91.8 | 0.0081 | 0.0088 |
| 71 | 8.72 | 486.0 | 0.14 | 3.01 | 47.58 | 93.1 | 93.0 | 0.0082 | 0.0088 |
| 72 | 8.57 | 483.1 | 0.15 | 2.95 | 47.30 | 93.1 | 92.7 | 0.0091 | 0.0094 |
| 73 | 8.70 | 486.0 | 0.16 | 2.91 | 47.81 | 93.1 | 92.7 | 0.0089 | 0.0089 |
| 74 | 8.80 | 484.4 | 0.14 | 2.92 | 47.97 | 93.1 | 92.4 | 0.0083 | 0.0089 |
| 75 | 8.75 | 485.0 | 0.16 | 2.89 | 47.95 | 93.0 | 92.4 | 0.0089 | 0.0090 |
| 76 | 8.48 | 486.0 | 0.16 | 3.00 | 47.15 | 93.0 | 93.3 | 0.0094 | 0.0091 |
| 77 | 8.66 | 482.4 | 0.16 | 2.94 | 47.42 | 93.0 | 92.5 | 0.0094 | 0.0095 |
| 78 | 8.78 | 485.1 | 0.19 | 2.66 | 48.84 | 93.0 | 91.2 | 0.0095 | 0.0091 |
| 79 | 8.57 | 483.4 | 0.12 | 3.00 | 47.38 | 93.0 | 92.8 | 0.0080 | 0.0090 |
| 80 | 8.53 | 485.1 | 0.14 | 3.13 | 46.89 | 93.0 | 93.6 | 0.0086 | 0.0090 |
| 81 | 8.54 | 484.1 | 0.14 | 3.02 | 47.30 | 93.0 | 93.0 | 0.0087 | 0.0092 |
| 82 | 8.74 | 485.0 | 0.18 | 2.90 | 47.83 | 92.9 | 92.5 | 0.0098 | 0.0092 |
| 83 | 8.80 | 484.0 | 0.17 | 3.01 | 47.37 | 92.9 | 93.0 | 0.0097 | 0.0093 |
| 84 | 8.35 | 484.4 | 0.13 | 2.89 | 46.52 | 92.9 | 93.4 | 0.0094 | 0.0093 |
| 85 | 7.92 | 485.1 | 0.13 | 2.96 | 46.68 | 92.9 | 93.3 | 0.0089 | 0.0092 |
| 86 | 8.71 | 486.0 | 0.16 | 3.00 | 47.61 | 92.9 | 93.0 | 0.0088 | 0.0089 |
| 87 | 8.74 | 483.1 | 0.14 | 2.79 | 47.76 | 92.9 | 92.1 | 0.0087 | 0.0092 |
| 88 | 8.65 | 483.1 | 0.14 | 2.98 | 47.60 | 92.9 | 92.6 | 0.0086 | 0.0092 |
| 89 | 8.58 | 483.3 | 0.16 | 2.93 | 47.60 | 92.9 | 92.5 | 0.0094 | 0.0095 |
| 90 | 8.68 | 483.3 | 0.13 | 2.95 | 46.35 | 92.9 | 93.7 | 0.0093 | 0.0092 |
| 91 | 8.57 | 483.1 | 0.12 | 2.92 | 47.41 | 92.9 | 92.6 | 0.0082 | 0.0091 |
| 92 | 8.66 | 482.6 | 0.15 | 2.85 | 47.85 | 92.8 | 92.0 | 0.0092 | 0.0095 |
| 93 | 8.66 | 484.3 | 0.15 | 2.99 | 47.61 | 92.8 | 92.7 | 0.0089 | 0.0092 |
| 94 | 8.40 | 483.7 | 0.14 | 3.03 | 47.17 | 92.8 | 93.0 | 0.0089 | 0.0093 |
| 95 | 8.58 | 486.0 | 0.15 | 2.94 | 47.86 | 92.8 | 92.6 | 0.0086 | 0.0089 |
| 96 | 8.42 | 484.8 | 0.17 | 3.01 | 47.14 | 92.8 | 93.1 | 0.0098 | 0.0093 |
| 97 | 8.72 | 482.3 | 0.17 | 2.90 | 47.86 | 92.8 | 92.1 | 0.0098 | 0.0096 |
| 98 | 8.65 | 485.0 | 0.16 | 2.96 | 47.70 | 92.8 | 92.7 | 0.0091 | 0.0091 |
| 99 | 8.39 | 484.3 | 0.16 | 2.86 | 47.83 | 92.8 | 92.2 | 0.0092 | 0.0093 |
| 100 | 8.57 | 486.0 | 0.15 | 2.90 | 48.02 | 92.8 | 92.4 | 0.0086 | 0.0089 |

| | | | | | | | | | | |
|---|---|---|---|---|---|---|---|---|---|---|
| 101 | 8.70 | 481.0 | 0.14 | 3.14 | 46.76 | 92.8 | 93.3 | 0.0092 | 0.0097 |
| 102 | 8.72 | 486.0 | 0.17 | 3.03 | 47.30 | 92.8 | 93.3 | 0.0096 | 0.0091 |
| 103 | 8.66 | 484.1 | 0.16 | 2.92 | 47.88 | 92.8 | 92.4 | 0.0090 | 0.0092 |
| 104 | 8.80 | 481.9 | 0.15 | 3.01 | 47.60 | 92.8 | 92.5 | 0.0090 | 0.0095 |
| 105 | 8.72 | 486.0 | 0.18 | 2.98 | 47.53 | 92.8 | 93.0 | 0.0098 | 0.0091 |
| 106 | 8.72 | 486.0 | 0.13 | 3.08 | 47.83 | 92.7 | 92.8 | 0.0077 | 0.0086 |
| 107 | 8.72 | 485.1 | 0.16 | 3.03 | 47.50 | 92.7 | 93.0 | 0.0091 | 0.0091 |
| 108 | 8.71 | 485.0 | 0.18 | 2.98 | 47.56 | 92.7 | 92.8 | 0.0100 | 0.0092 |
| 109 | 8.64 | 481.5 | 0.15 | 3.04 | 47.26 | 92.7 | 92.7 | 0.0094 | 0.0097 |
| 110 | 8.80 | 486.0 | 0.18 | 2.95 | 47.93 | 92.7 | 92.7 | 0.0092 | 0.0090 |
| 111 | 8.75 | 485.9 | 0.18 | 2.77 | 48.70 | 92.7 | 91.7 | 0.0092 | 0.0090 |
| 112 | 8.58 | 484.9 | 0.19 | 2.81 | 48.32 | 92.7 | 91.8 | 0.0097 | 0.0093 |
| 113 | 8.74 | 485.9 | 0.17 | 2.91 | 48.18 | 92.6 | 92.4 | 0.0089 | 0.0089 |
| 114 | 8.66 | 486.0 | 0.13 | 3.01 | 48.17 | 92.6 | 92.5 | 0.0076 | 0.0086 |
| 115 | 8.80 | 484.3 | 0.17 | 2.96 | 47.94 | 92.6 | 92.5 | 0.0093 | 0.0092 |
| 116 | 8.57 | 486.0 | 0.19 | 2.92 | 47.81 | 92.6 | 92.6 | 0.0100 | 0.0091 |
| 117 | 8.58 | 483.7 | 0.13 | 3.06 | 47.66 | 92.6 | 92.7 | 0.0080 | 0.0090 |
| 118 | 8.68 | 484.7 | 0.12 | 3.03 | 48.14 | 92.6 | 92.4 | 0.0076 | 0.0087 |
| 119 | 8.66 | 482.6 | 0.15 | 2.85 | 48.28 | 92.6 | 91.6 | 0.0090 | 0.0094 |
| 120 | 8.60 | 485.0 | 0.16 | 2.89 | 48.25 | 92.5 | 92.1 | 0.0088 | 0.0091 |
| 121 | 8.80 | 486.0 | 0.16 | 2.95 | 48.35 | 92.5 | 92.3 | 0.0084 | 0.0088 |
| 122 | 8.80 | 484.9 | 0.18 | 2.80 | 48.80 | 92.4 | 91.5 | 0.0092 | 0.0091 |
| 123 | 8.59 | 485.1 | 0.14 | 2.88 | 48.51 | 92.4 | 91.9 | 0.0081 | 0.0088 |
| 124 | 8.80 | 483.4 | 0.16 | 2.98 | 48.11 | 92.4 | 92.2 | 0.0088 | 0.0092 |
| 125 | 8.55 | 484.9 | 0.16 | 3.02 | 47.62 | 92.4 | 92.8 | 0.0092 | 0.0092 |
| 126 | 8.73 | 485.1 | 0.19 | 2.96 | 47.89 | 92.4 | 92.6 | 0.0099 | 0.0092 |
| 127 | 8.60 | 485.0 | 0.17 | 2.93 | 48.03 | 92.4 | 92.4 | 0.0093 | 0.0092 |
| 128 | 8.80 | 485.0 | 0.17 | 3.03 | 47.85 | 92.4 | 92.7 | 0.0089 | 0.0091 |
| 129 | 8.72 | 482.5 | 0.17 | 3.02 | 47.72 | 92.4 | 92.5 | 0.0093 | 0.0095 |
| 130 | 8.56 | 483.3 | 0.13 | 2.99 | 48.03 | 92.4 | 92.2 | 0.0082 | 0.0091 |
| 131 | 8.50 | 482.4 | 0.09 | 3.15 | 46.68 | 92.3 | 93.4 | 0.0076 | 0.0089 |
| 132 | 8.51 | 481.5 | 0.14 | 3.01 | 46.93 | 92.4 | 92.9 | 0.0096 | 0.0098 |
| 133 | 8.80 | 486.0 | 0.17 | 3.05 | 47.69 | 92.3 | 93.0 | 0.0091 | 0.0090 |
| 134 | 8.70 | 483.1 | 0.14 | 3.07 | 47.88 | 92.3 | 92.5 | 0.0082 | 0.0091 |
| 135 | 8.73 | 486.0 | 0.14 | 2.95 | 48.71 | 92.3 | 91.9 | 0.0078 | 0.0086 |
| 136 | 8.50 | 485.0 | 0.14 | 3.04 | 47.87 | 92.2 | 92.6 | 0.0083 | 0.0090 |
| 137 | 8.69 | 483.7 | 0.14 | 3.06 | 48.03 | 92.2 | 92.4 | 0.0081 | 0.0090 |
| 138 | 8.80 | 485.0 | 0.15 | 2.96 | 48.53 | 92.2 | 92.1 | 0.0082 | 0.0089 |
| 139 | 8.69 | 483.6 | 0.12 | 3.01 | 48.45 | 92.2 | 91.9 | 0.0076 | 0.0088 |
| 140 | 8.65 | 484.0 | 0.17 | 2.93 | 48.25 | 92.2 | 92.0 | 0.0091 | 0.0093 |
| 141 | 8.80 | 485.1 | 0.17 | 2.96 | 48.35 | 92.2 | 92.2 | 0.0087 | 0.0090 |
| 142 | 8.57 | 482.8 | 0.14 | 2.92 | 48.38 | 92.2 | 91.7 | 0.0085 | 0.0093 |
| 143 | 8.80 | 484.4 | 0.16 | 2.92 | 48.58 | 92.2 | 91.9 | 0.0086 | 0.0090 |
| 144 | 8.71 | 483.1 | 0.15 | 2.94 | 47.64 | 93.1 | 85.0 | 0.0089 | 0.0463 |
| 145 | 8.72 | 485.0 | 0.14 | 3.05 | 48.25 | 92.1 | 92.4 | 0.0080 | 0.0089 |
| 146 | 8.63 | 484.2 | 0.16 | 3.04 | 47.85 | 92.1 | 92.6 | 0.0088 | 0.0092 |
| 147 | 8.72 | 481.9 | 0.14 | 2.89 | 47.18 | 92.1 | 92.6 | 0.0093 | 0.0095 |
| 148 | 8.51 | 483.7 | 0.19 | 2.79 | 48.79 | 92.1 | 91.1 | 0.0099 | 0.0095 |
| 149 | 7.57 | 485.2 | 0.13 | 2.98 | 47.41 | 92.0 | 92.5 | 0.0089 | 0.0094 |
| 150 | 8.73 | 483.5 | 0.19 | 2.95 | 48.18 | 92.0 | 92.1 | 0.0099 | 0.0095 |

**L**: reactor length (m); **T**: reactor temperature (K); **SR**: splitter ratio; **P**: system pressure (MPa); **CO**: CO molar flow rate (mol/s)